\documentclass[aps,prb,twocolumn,nobibnotes,superscriptaddress]{revtex4-2}  
\usepackage{dcolumn}
\usepackage{graphicx}
\usepackage{color}
\usepackage{amssymb}
\usepackage{amsmath}
\usepackage{bm}
\usepackage{bbm}
\usepackage[colorlinks=true,linktocpage=true,breaklinks=true]{hyperref}
\usepackage{multirow}
\usepackage{braket}
\usepackage{appendix}
\usepackage{mathtools}

\begin{document}
\newcommand{\aurelien}[1]{\textcolor{green}{Aurélien: #1}}
\newcommand{\diego}[1]{\textcolor{blue}{Diego: #1}}
\title{Spin and orbital-to-charge conversion in noncentrosymmetric materials:\\ Hall versus Rashba-Edelstein effects}

\begin{abstract}
We investigate spin- and orbital-to-charge conversion phenomena in nonmagnetic materials with broken inversion symmetry, treating the contributions from the Hall effect and the Rashba–Edelstein effect on an equal footing. We develop a general formalism for this interconversion based on macroscopic observables (susceptibility, conductivity, conversion efficiencies). The theory is validated through a case study of ferroelectric $\alpha$-GeTe, where we find that the effective Rashba parameter obtained is smaller than previously reported values for the same material. Incorporating these parameters into a drift–diffusion model, we show that the generated charge current is primarily governed by the Rashba–Edelstein effect, rather than by the spin or orbital Hall effects.\end{abstract}
\author{Diego Garc\'ia Ovalle}
\email{diego.garcia.ovalle@usherbrooke.ca}
\affiliation{Département de physique et Institut quantique, Université de Sherbrooke, Sherbrooke J1K 2R1 QC, Canada}
\author{Aur\'{e}lien Manchon}
\email{aurelien.manchon@univ-amu.fr}
\affiliation{Aix-Marseille Universit\'e, CNRS, CINaM, Marseille, France}

\date{\today}
\maketitle

\section{Introduction}

The optimization of the conversion between charge and spin currents is a central goal of modern spintronics. This can be done, among other ways, using a wide variety of quantum materials and heterostructures \cite{Han2018}, which must display a sizable spin-orbit interaction and/or a non-trivial spin texture (via noncollinear magnetism, for instance). The spin-charge interconversion usually occurs through two classes of mechanisms: the spin Hall effect (SHE) and the spin Rashba-Edelstein effect (SREE). The SHE features the generation of a pure spin current $\mathbf{J}_s$, transverse to the injected charge current $\mathbf{J}_c$ \cite{Dyakonov1971,Sinova2015}. The SREE features the generation of a spin density ${\bf S}$, induced by a charge current $\mathbf{J}_c$ \cite{Edelstein1990}, and only exists in systems lacking inversion symmetry \cite{Bihlmayer2022}. To assess the interconversion efficiency of a given spin-charge converter, the conventional protocol involves depositing a ferromagnet in proximity to the converter and using it as a source or drain of spin current. 

In most experiments, the converter is a (centrosymmetric) heavy metal such as Pt, W, or Ta \cite{Mosendz2010,Ando2010,Mosendz2010b,Pai2012,Jiao2013,Wang2014,RojasSanchez2014,Isasa2015,Yu2018}, and the observed signal is generally attributed to the SHE occurring in the bulk of the material. In this context, the SREE, which naturally arises at the heavy-metal/ferromagnet interface, is often neglected — an approximation justified by the high conductivity of the heavy metal \cite{Manchon2024}.
Other experiments instead exploit the large SREE present at interfaces or surfaces with strong spin–orbit coupling, such as the surfaces of topological insulators (e.g., Bi$_2$Se$_3$ or $\alpha$-Sn) \cite{Zhang2016,RojasSanchez2016}, oxide two-dimensional electron gases \cite{Lesne2016,Wang2017b,Vaz2019,Noel2020,Arche2022,Varotto2022}, or at the interface between certain metallic alloys, like Bi/Ag, and the ferromagnet \cite{RojasSanchez2013b,Yue2018,Cheng2022}. In these systems, spin–charge interconversion is usually attributed entirely to the interfacial SREE \cite{Chen2015c}, since any bulk SHE contribution is negligible or absent. However, this simplified picture — assigning spin–charge conversion exclusively to either the SHE or the SREE — has recently been challenged. Comparative studies of structures such as (heavy metal/Cu)/YIG versus (Cu/heavy metal)/YIG \cite{Yu2020}, and Pt/NiFe versus Bi$_2$Te$_3$/NiFe \cite{Shi2025}, provide evidence for the coexistence of both mechanisms, SHE and interfacial SREE.

\begin{figure}[ht!]
    \vspace{-0.4cm}		
	\includegraphics[width=1\linewidth]{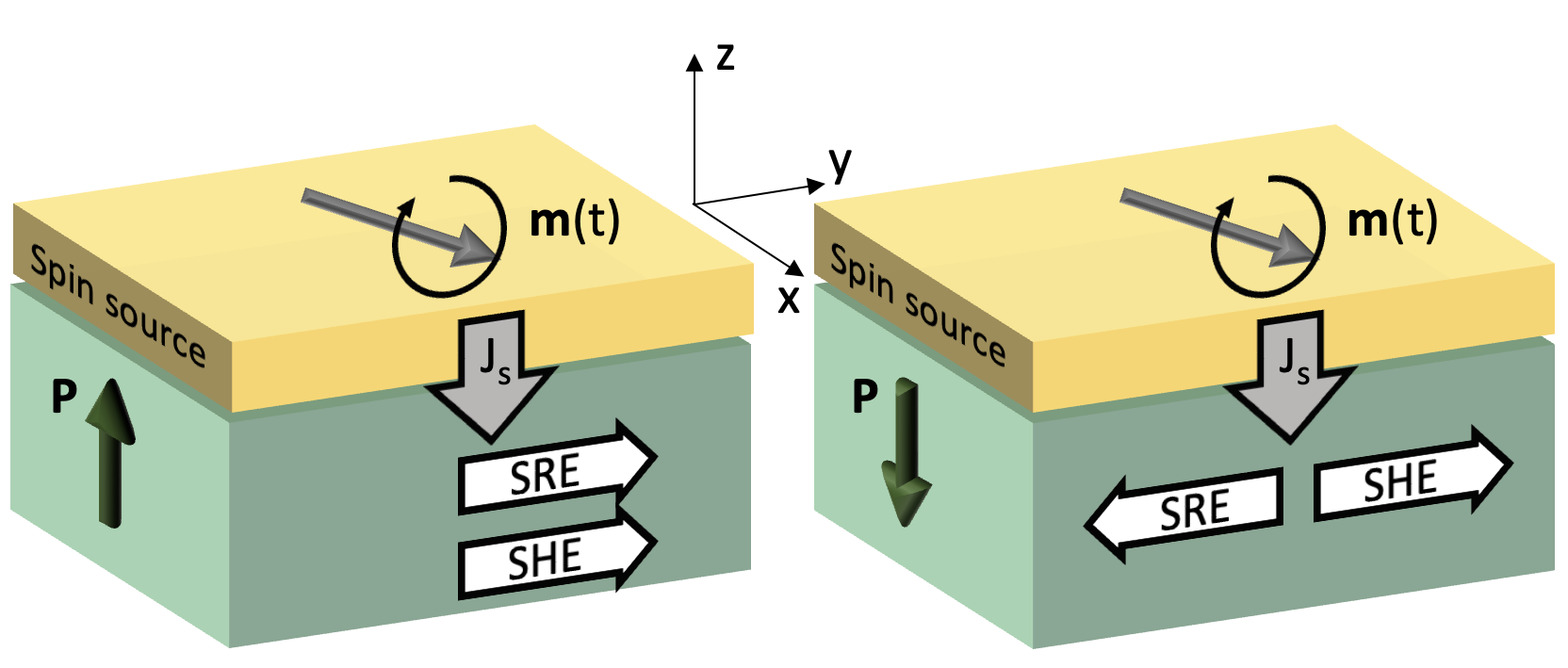}
    \caption{(Color online) Sketch of the spin-to-charge conversion process taking place in a conductive ferroelectric/ferromagnet bilayer. The ferromagnet pumps a spin current (grey arrow) into the ferroelectric, which is then converted into a charge current (white arrows) through SHE and SREE. }
	\label{fig0}
	\end{figure}

The situation is even more dramatic in {\em noncentrosymmetric} materials such as WTe$_2$ \cite{Sharma2019}, BiTeI \cite{Ishizaka2011}, $\alpha$-GeTe \cite{Varotto2021}, or Te \cite{Calavalle2022}, where both SREE and SHE are present in the material's bulk. A straightforward way to distinguish between the SREE and SHE in these systems is to reverse the material's polarity, since only the SREE is expected to change sign, as depicted in Fig. \ref{fig0}. While this is challenging in practice, it can be achieved experimentally in ferroelectric materials. In fact, because the Rashba-Edelstein effect is tied to the chirality of the spin texture, reversing the ferroelectric polarization directly switches this spin texture \cite{Bahramy2011,DiSante2013,Jafari2022,Cai2022,Tensin2023}. Such a change of sign upon ferroelectric switching was recently reported in $\alpha$-GeTe \cite{Varotto2021} and interpreted by considering that the polarization reversal leads to two dissimilar crystal structures, resulting in distinct SHE signatures.

This experiment motivated us to develop a theory that combines diffusive transport, linear response theory, and first-principles calculations to describe spin-to-charge conversion in such bilayers, treating the spin Rashba–Edelstein and spin Hall effects on an equal footing. Considering the current interest in orbital-charge interconversion effects \cite{Go2018,Choi2023}, we also extend our theory to the atomic orbital Hall effect \cite{Jo2018,Pezo2022} (OHE) and orbital Rashba-Edelstein effect \cite{Go2017} (OREE). Our paper is organized as follows: In Section \ref{II}, we derive the drift-diffusion equations for coexisting SHE and SREE and determine the conversion coefficients. In Section \ref{III}, the theory is applied to $\alpha$-GeTe and extended to orbital-to-charge conversion. Finally, in Section \ref{IV} we provide the conclusions and further discussion of the work. 

\section{General theory}\label{II}

\subsection{Drift diffusion model}

We consider a bilayer composed of a ferromagnetic metal (the spin source) deposited on top of a ferroelectric conductor (here, $\alpha$-GeTe), as depicted in Fig. \ref{fig0}. Excited at magnetic resonance, the ferromagnet pumps a spin current ${\bf J}_s^{(0)}$ (in units of ($\hbar/2e$) A/cm$^2$) into the ferroelectric, inducing a spin chemical potential ${\bm\mu}_s$ (in V) \cite{Tserkovnyak2002b,Brataas2002}, which, in turn, produces a charge current via SHE and SREE. As mentioned above, SHE is invariant under reversing the ferroelectric polarization, while SREE switches sign. The charge current (in A/cm$^2$) produced via (inverse) SHE is directly connected to ${\bm\mu}_s$ by \cite{Shchelushkin2005b}
\begin{eqnarray}\label{eq:she}
{\bf J}_c^{{\rm SHE}}&=&\sigma_{\rm H}{\bm\nabla}\times{\bm\mu}_s,
\end{eqnarray}
where $\sigma_{\rm H}$ is the (inverse) spin Hall conductivity (in $\Omega^{-1}\cdot$m$^{-1}$). For the SREE, we notice that in the case of $\alpha$-GeTe, the spin-orbit coupling takes the form ${\cal H}_{\rm R}=\alpha_R ({\bf k}\times {\bf z})\cdot{\bm\sigma}$, where $\alpha_R$ is the Rashba parameter (in eV$\cdot$m) \cite{Picozzi2014,Liebmann2016}. We immediately find that the charge current reads
\begin{eqnarray}\label{eq:{SREE}}
{\bf J}_c^{\rm SREE}&=&-\frac{e\alpha_R}{\hbar}{\bf z}\times{\bf S}=-\frac{e^2}{\hbar}{\cal N}\alpha_R{\bf z}\times{\bm\mu}_s.
\end{eqnarray}
Here, the spin accumulation is given by ${\bf S}=e{\cal N}{\bm\mu}_s$ (in units of 1/m$^3$), being ${\cal N}$ the density of states. In a multiorbital system, $\alpha_R$ must be replaced by an effective parameter average over the band structure, $\bar{\alpha}_R$ (see below). Then, as depicted in Fig. \ref{fig0}, the spin density is aligned along ${\bf x}$, and both SHE and SREE produce a charge current along ${\bf y}$. The profile of the spin chemical potential is determined using standard drift-diffusion theory,
\begin{align}
    \mu_s(z)=\frac{J_s^{(0)}/G^{\uparrow\downarrow}}{1+\frac{\sigma_{\rm N}\tanh \frac{d_{\rm N}}{\lambda}}{2\lambda G^{\uparrow\downarrow}}}\left(\cosh\frac{z}{\lambda}+\tanh\frac{d_{\rm N}}{\lambda}\sinh\frac{z}{\lambda}\right).\label{drift5}
\end{align}
Here, $\lambda$ is the spin relaxation length, $G^{\uparrow\downarrow}$ is the interfacial spin-mixing conductance, and $d_{\rm N}$ is the thickness of the ferroelectric conductor. In Eq. \eqref{drift5}, the spin backflow is responsible for the renormalization factor, and the interfacial spin memory loss was neglected. Accounting for it leads to a reduction of the interfacial spin density $\mu_s(0)$ effectively induced by the spin current $J_s^{(0)}$, but it does not affect the overall thickness dependence (see, e.g., Ref. \cite{RojasSanchez2014}). The charge current induced by SHE reads
\begin{align}
    J_c^{\rm SHE}=\frac{\sigma_{\rm H}}{d_{\rm N}}\int_{-d_{\rm N}}^0\partial_z\mu_s dz=-\frac{J_s^{(0)}}{G^{\uparrow\downarrow}}\frac{\sigma_{\rm H}}{d_{\rm N}}\frac{1-\cosh^{-1}\frac{d_{\rm N}}{\lambda}}{1+\frac{\sigma_{\rm N}\tanh \frac{d_{\rm N}}{\lambda}}{2\lambda G^{\uparrow\downarrow}}}.\label{drift6}
\end{align}
In addition, the current produced by SREE reads \cite{Manchon2024}
\begin{eqnarray}
    J_c^{\rm SREE}&=&-\frac{e^2\mathcal{N}\bar{\alpha}_R}{\hbar d_{\rm N}}\int_{-d_{\rm N}}^0{\mu}_sdz,\nonumber\\
    &=&-\frac{J_s^{(0)}}{G^{\uparrow\downarrow}}\frac{e^2\mathcal{N}\bar{\alpha}_R}{\hbar}\frac{\lambda}{d_{\rm N}}\frac{\tanh \frac{d_{\rm N}}{\lambda}}{1+\frac{\sigma_{\rm N}\tanh \frac{d_{\rm N}}{\lambda}}{2\lambda G^{\uparrow\downarrow}}},\label{drift7}
\end{eqnarray}

\begin{figure}[ht!]
    \vspace{-0.4cm}		
	\includegraphics[width=0.8\linewidth]{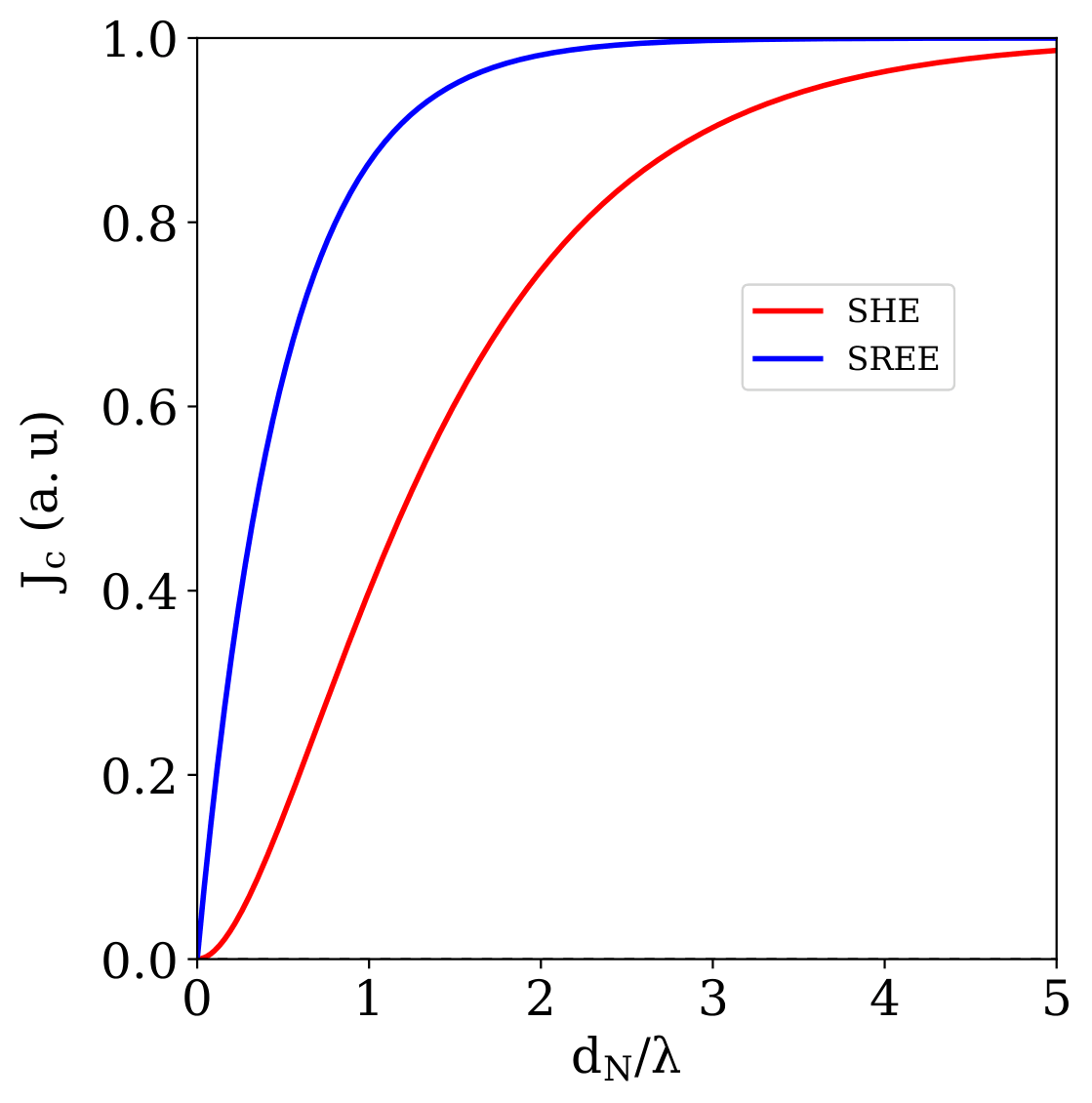}
    \caption{(Color online) Thickness dependence of the charge current induced by inverse SHE (red) or SREE (blue), based on Eqs. \eqref{drift6}-\eqref{drift7}. The currents are normalized to their value at infinite thickness, and we set $\sigma_{\rm N}/G^{\uparrow\downarrow}\lambda$=1.}
	\label{fig0b}
	\end{figure}

The dependence of the charge current induced by inverse SHE or SREE as a function of the normal metal thickness is reported in Fig. \ref{fig0b}. The two effects exhibit a different behavior at thin thicknesses, when $d_{\rm N}<\lambda$, which could, in principle, be used to distinguish them in experiments. Nonetheless, this regime is difficult to study experimentally, as roughness tends to degrade at such ultrathin thicknesses. In addition, it is reasonable to expect that both effects generally coexist, which makes the separation challenging.

To assess the relative magnitude of the two effects, we define the ratio
\begin{eqnarray}
    \Xi=\frac{J_c^{\rm SHE}}{J_c^{\rm SREE}}=\left(\frac{\hbar\sigma_{\rm H}}{e^2\mathcal{N}\bar{\alpha}_R\lambda}\right)\left(\frac{\cosh\frac{d_{\rm N}}{\lambda}-1}{\sinh \frac{d_{\rm N}}{\lambda}}\right).\label{drift8}
\end{eqnarray}
Notice that this theory can be easily extended to consider orbital-charge interconversion, as the process of orbital pumping is physically very similar to spin pumping \cite{Han2025,Go2025,Pezo2025}. To obtain a quantitative estimate of the spin- and orbital-to-charge conversion efficiencies, we need to compute $\sigma_{\rm H}$, ${\cal N}$, and $\bar{\alpha}_R$ directly from first principles.

\subsection{Spin-to-charge conversion coefficients from macroscopic observables} \label{Theory}

The electrical generation of a spin current (SHE) or a spin accumulation (SREE) can be readily computed using perturbation theory, e.g., the Kubo formula, because the electric field couples directly to the position operator (or, equivalently, the vector potential to the velocity operator). In the case of a nonmagnetic converter, intrinsic SHE involves only interband transitions, governed by the Fermi sea, whereas the SREE involves only intraband transitions, governed by the Fermi surface, as dictated by their different parity under time-reversal \cite{Bonbien2020}. The reciprocal (or inverse) effects, Eqs. \eqref{eq:she}-\eqref{eq:{SREE}}, can be obtained from the Onsager reciprocity theorem \cite{Onsager1931a,Onsager1931b}, which connects {\em generalized forces} (or, in other words, gradients of potentials) to {\em generalized currents}.

\subsubsection{Inverse spin Hall effect}

Since the {\em direct} SHE is driven by the charge chemical potential gradient and the {\em inverse} SHE is driven by the spin chemical potential gradient, the spin-to-charge and charge-to-spin conductivities are simply equal to each other. The direct SHE is defined $J_{s,\gamma}^\alpha=\sigma_{\alpha\beta}^{\gamma}\rm E_\beta$, where $J_{s,\gamma}^\alpha$ is the spin Hall current and $\rm E_\beta$ is the electric field. The intrinsic spin conductivity $\sigma_{\alpha\beta}^{\gamma}$ (in units of $(\hbar/2e)$ $\Omega^{-1}\cdot$m$^{-1}$) reads \cite{Qiao2018}
\begin{align}
\sigma_{\alpha\beta}^{\gamma}=-\frac{e}{\hbar}\int_{BZ}\frac{d^3{\bf k}}{(2\pi)^3}\sum_n\Omega_{n\mathbf{k},\alpha\beta}^{\gamma} f_{n\mathbf{k}},\label{scc1}
\end{align}
where $f_{n\mathbf{k}}$ is the equilibrium Fermi distribution function, and the spin Berry curvature reads
\begin{align}
    \Omega_{n\mathbf{k},\alpha\beta}^{\gamma}&=-2\hbar^2\operatorname{Im}\sum_{m\not=n}\frac{\bra{u_{n\mathbf{k}}}\hat{v}_{\mathbf{k},\alpha}^{\gamma}\ket{u_{m\mathbf{k}}}\bra{u_{m\mathbf{k}}}\hat{v}_{\mathbf{k},\beta}\ket{u_{n\mathbf{k}}}}{(\epsilon_{n\mathbf{k}}-\epsilon_{m\mathbf{k}})^2}\label{scc2}.
\end{align}
Here, given a periodic Hamiltonian $\mathcal{H}_\mathbf{k}$ with eigenstates $\ket{u_{n\mathbf{k}}}$ such that $\mathcal{H}_\mathbf{k}\ket{u_{n\mathbf{k}}}=\epsilon_{n\mathbf{k}}\ket{u_{n\mathbf{k}}}$, we take $\hat{\bm v}_{\mathbf{k}}=\hbar^{-1}\partial_\mathbf{k}\mathcal{H}_\mathbf{k}$ and $\hat{\bm v}_{\mathbf{k}}^{\gamma}=\frac{\hbar}{4}\left\{\hat{s}_\gamma,\hat{\bm v}_{\mathbf{k}}\right\}$, with $\hat{\bf s}$ the vector of Pauli spin-1/2 matrices. Unless stated otherwise, we consider all the calculations at zero temperature limit, where $f_{n\mathbf{k}}\to \Theta(\epsilon_F-\epsilon_{n\mathbf{k}})$, being $\epsilon_{\rm F}$ the Fermi energy and $\Theta(x)$ the Heaviside step function. The longitudinal conductivity can be accounted for through
\begin{align}
    \sigma_{xx}&=-e^2\tau\int_{BZ}\frac{d^3{\bf k}}{(2\pi)^3}(v_{n\mathbf{k},x})^2\partial_{\epsilon_{n\mathbf{k}}}f_{n\mathbf{k}}, \label{scc4}
\end{align}
where $\tau$ is the constant momentum scattering time. Applying Onsager reciprocity to Eq. \eqref{eq:she}, one simply gets $\sigma_{\rm H}=(2e/\hbar)\sigma_{xy}^z$.

\subsubsection{Inverse Rashba-Edelstein effect} \label{REE}

Let us now consider the direct SREE and inverse SREE, expressed respectively as 
\begin{eqnarray}
{\bf S}=\eta_{c\rightarrow s}{\bf z}\times {\bf J}_c,\label{eq:cs}\\
{\bf J}_c=\eta_{s\rightarrow c}{\bf z}\times {\bf S}.\label{eq:sc}
\end{eqnarray}
The direct SREE coefficient, $\eta_{c\rightarrow s}$ (in units of A$^{-1}\cdot$m$^{-1}$), is given by the spin-current correlation function, readily computed using the Kubo formula in the weak disorder limit and constant relaxation time approximation \cite{Zhong2016}

\begin{align}
\eta_{c\rightarrow s}&=&-\frac{e\tau}{\sigma_{xx}}\int_{BZ}\frac{d^3{\bf k}}{(2\pi)^3}\sum_ns_{n\mathbf{k},x}v_{n\mathbf{k},y}\partial_{\epsilon_{n\mathbf{k}}}f_{n\mathbf{k}},\label{scc5}
\end{align}
The reciprocal effect, $\eta_{s\rightarrow c}$ (in units of A$\cdot$m), is proportional to the current-spin correlation function, but cannot be computed by applying Eq. \eqref{scc5} directly. On one hand, since the spin naturally couples to a magnetic field via $\mathcal{H}_{int}=\mu_B\mathbf{B}\cdot\mathbf{S}$, the charge current induced by a time-dependent magnetic field reads $\mathbf{J}_c=\mu_B\chi_s\dot{\mathbf{B}}$, with $\chi_s=\sigma_{xx}\eta_{c\rightarrow s}$ (in of V$^{-1}\cdot$m$^{-2}$). On the other hand, the spin density induced by a time-dependent magnetic field is $\mathbf{S}=\Sigma_B\dot{\mathbf{B}}$, where $\Sigma_B$ is the spin-spin correlation tensor, or spin susceptibility, at the Fermi energy,  and reads

\begin{align}
    \Sigma_{B}^{\alpha\beta}&=\tau\mu_B\int_{BZ}\frac{d^3{\bf k}}{(2\pi)^3}s_{n\mathbf{k},\alpha}s_{n\mathbf{k},\beta}\partial_{\epsilon_{n\mathbf{k}}}f_{n\mathbf{k}}.\label{scc7}
\end{align}

Therefore, assuming that the spin susceptibility is diagonal (weak spin-orbit coupling), the spin-to-charge conversion corresponds to the ratio between Eq. \eqref{scc5} and Eq. \eqref{scc7}

\begin{eqnarray}
    \eta_{s\to c}=\mu_B\frac{\chi_s}{\Sigma_B}=\mu_B\frac{\sigma_{xx}}{\Sigma_B}\eta_{c\to s},\label{scc8}
\end{eqnarray}
that is independent of $\tau$. Clearly, direct and inverse SREE are not Onsager-reciprocal. In the next section, we compute these coefficients in the case of the ferroelectric phase of bulk $\alpha$-GeTe, comparing the effects from both spin and atomic orbital angular momentum.

\section{Application to Ferroelectric $\alpha$-GeTe}\label{III}
\subsection{Structure}
\begin{figure}[ht!]		
	\includegraphics[width=\columnwidth]{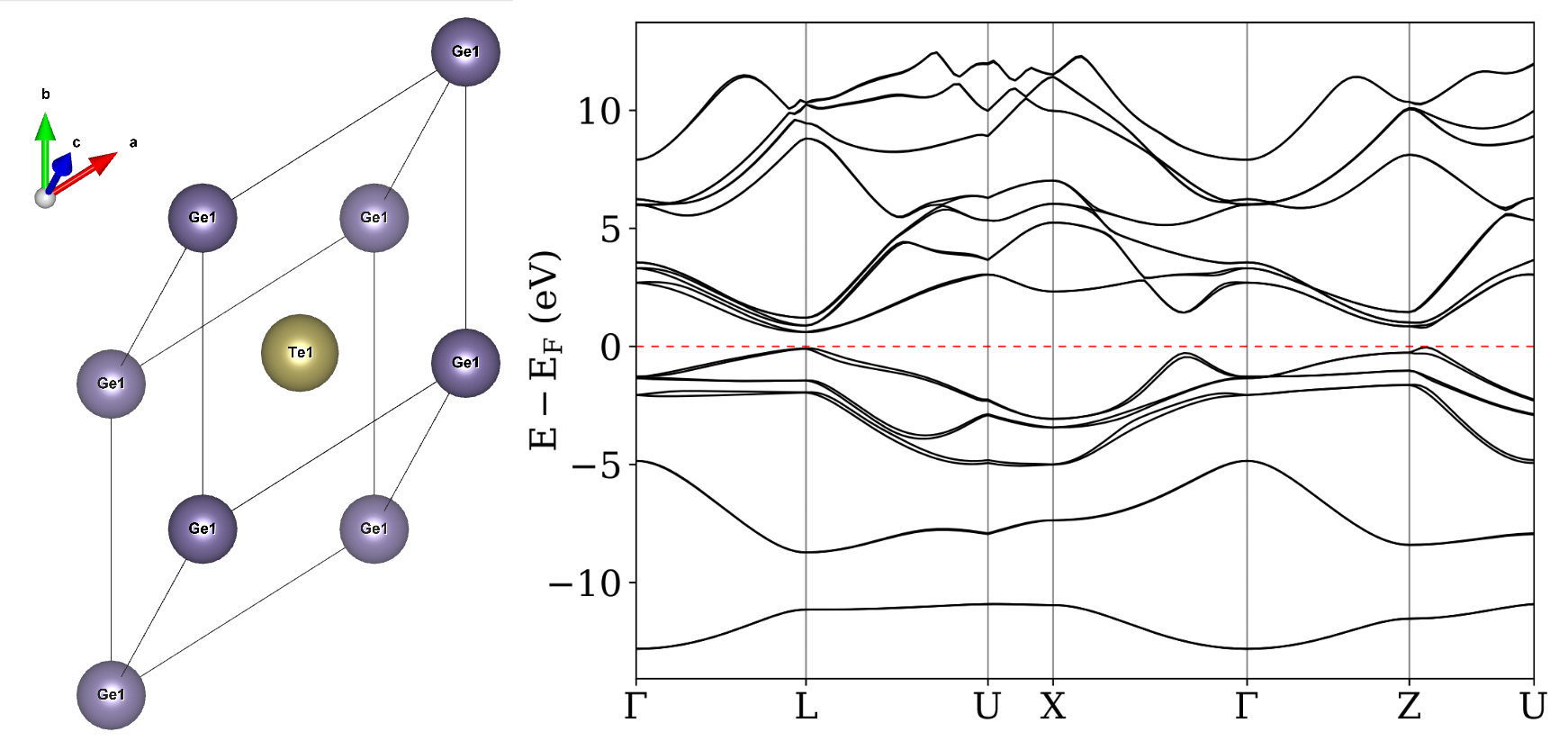}
    \caption{(Color online) Crystal structure and band structure with spin-orbit coupling of ferroelectric GeTe. The displacement of the Te atom with respect to the inversion center produces a ferroelectric polarization along the (111) direction, which we aligned with the $\mathbf{z}$ direction for simplicity. The former is obtained using VESTA \cite{Vesta2011}. We fix the Fermi level at the valence band maximum. The Rashba spin-splitting is visible close to the gap at the Z-point (see Fig. 1A in Ref. \cite{Picozzi2014})}. 
	\label{fig1}
	\end{figure}

We obtain the electronic band structure of $\alpha$-GeTe, which belongs to the C$_{3\mathrm{v}}$ (R3m) space group, applying density functional theory and the plane-wave basis set method implemented in QUANTUM ESPRESSO \cite{Giannozzi2009,Giannozzi2017}. We use ultra-soft pseudopotentials with nonlinear core correction and the PBE exchange-correlation functional \cite{Perdew1996}. The energy cutoff for the wavefunctions and the charge density are set as 70 Ry and 280 Ry, respectively. We perform the calculations in a grid of 20$\times$20$\times$20, taking into account the lattice parameters $a=4.373$ \AA \ and $\cos\gamma\simeq 0.533$. The band structure with spin-orbit interaction and the crystal structure are depicted in Fig. \ref{fig1}. Next, we build the tight-binding Hamiltonian from the projected Wannier functions on the $s$ and $p$ orbitals of Ge and Te, respectively, using the Wannier90 package \cite{Pizzi2020}. Then, we compute the transport coefficients presented in section \ref{Theory} using the WannierBerri package\cite{Tsirkin2021} for the spin contributions and a home-made implementation for the case of the atomic angular momentum. We set a value of $\tau=10^{-13}$ s in the simulations whenever it is necessary. 

\subsection{Spin-to-charge interconversion}

Let us first consider the symmetry restrictions on the charge-to-spin and orbital conversion tensor predicted from the Kubo formula. In this case, the symmetry analysis of the C$_{3\rm v}$ point group indicates that the spin (or orbital) Hall conductivity tensor takes the form \cite{Zhang2020,Wang2020} 

\begin{align}
  \hat{\sigma}^x=\begin{pmatrix}
\sigma_{xx}^x & 0 & 0\\
0 & -\sigma_{xx}^x & \sigma_{yz}^x\\
0 & -\sigma_{zx}^y & 0
\end{pmatrix}&, \quad \hat{\sigma}^y=\begin{pmatrix}
0 & -\sigma_{xx}^x & -\sigma_{yz}^x\\
-\sigma_{xx}^x & 0 & 0\\
\sigma_{zx}^y & 0 & 0
\end{pmatrix},\nonumber\\
\hat{\sigma}^z&=\begin{pmatrix}
0 & \sigma_{xy}^z & 0\\
-\sigma_{xy}^z & 0 & 0\\
0 & 0 & 0
\end{pmatrix}.\label{gete1}
\end{align}
and the charge conductivity tensor is 
\begin{align}
    \hat{\sigma}=\begin{pmatrix}
\sigma_{xx} & 0 & 0\\
0 & \sigma_{xx} & 0\\
0 & 0 & \sigma_{zz}
\end{pmatrix}.\label{gete2}
\end{align}

\begin{figure}[ht!]
	\includegraphics[width=\columnwidth]{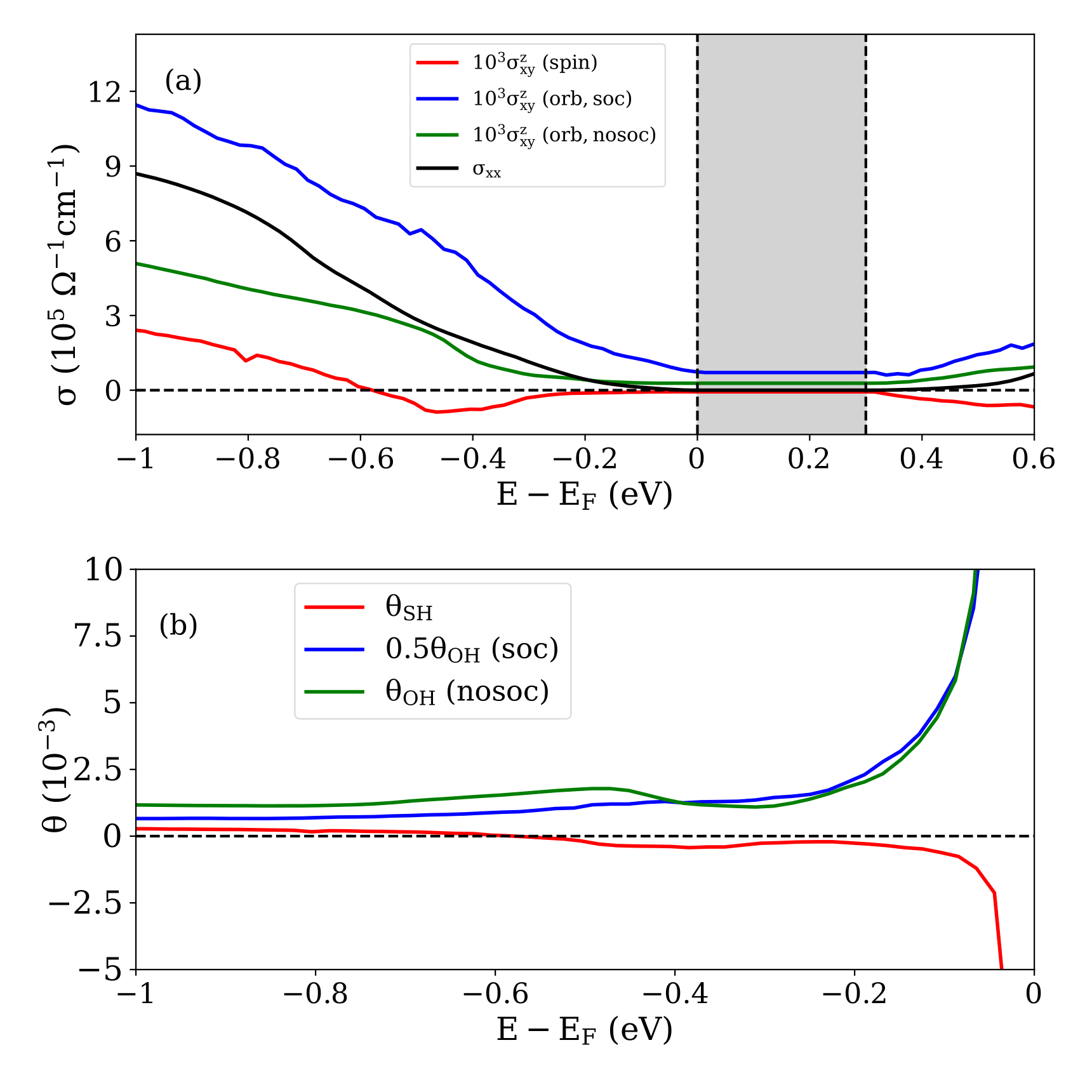}
    \caption{(Color online) (a) Longitudinal conductivity (black line), and one of the spin Hall conductivity coefficients (red line), as a function of the energy. We also show the orbital Hall conductivity with (blue) and without (green) spin-orbit coupling. (b) Spin Hall angle (red) and orbital Hall angle with (blue) and without (green) spin-orbit coupling.}
	\label{fig2}
	\end{figure}

Since the magnitude of the spin Hall angle is not dramatically sensitive to the component of the spin Hall conductivity examined, we illustrate the spin Hall angle associated with $\sigma_{xy}^z$ as a function of the energy. The longitudinal conductivity $\sigma_{xx}$, spin Hall, and orbital Hall conductivities (the latter is discussed in the next section) are displayed in Fig. \ref{fig2}(a). Our results are consistent with previous calculations \cite{Wang2020}, including the change of sign of the spin Hall conductivity at $\rm E-\rm E_F=-0.6$ eV, approximately. The corresponding spin Hall angle in the valence band is shown in Fig. \ref{fig2}(b), reaching values of the order of $10^{-2}$ close to the gap and decreasing progressively away from it. Notice that the magnitude of the spin Hall angle is strongly damped by the large longitudinal conductivity close to the valence band maximum, which scales as $10^3 \Omega^{-1}m^{-1}$. This spin Hall angle can reach one order of magnitude larger than in $n$-doped Ge at room temperature \cite{RojasSanchez2013}. 

\begin{figure}[ht!]
	\includegraphics[width=\columnwidth]{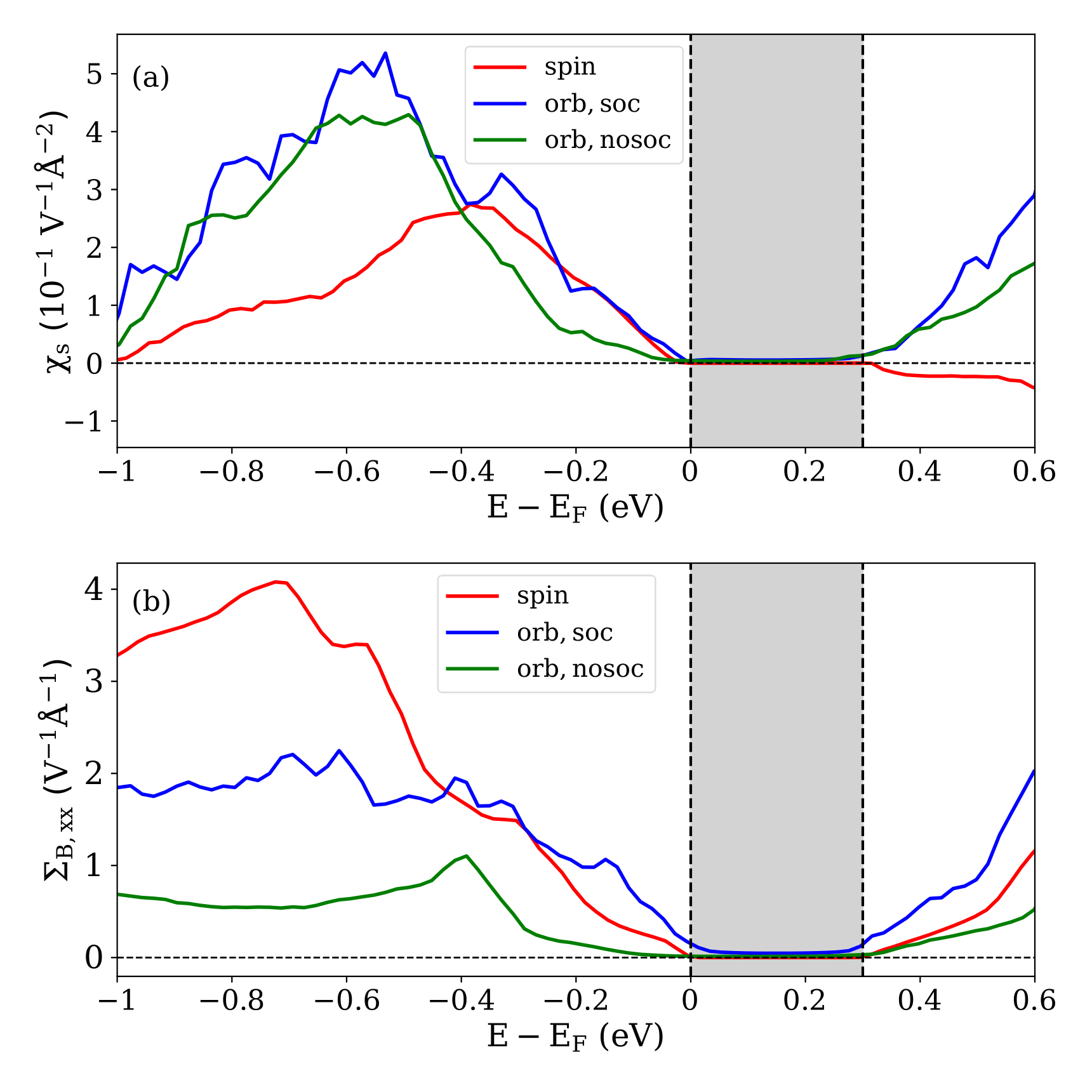}
    \caption{(Color online) (a) The SREE (red) and OREE coefficients with (blue) and without spin-orbit coupling (green) as a function of the energy. (b) Corresponding magnetic susceptibility as a function of the energy.}
	\label{fig3}
	\end{figure} 

The ferroelectric polarization $\mathbf{P}=\mathrm{P}_z\mathbf{z}$ in $\alpha-$GeTe arises from the displacement of the Te atom, producing a Rashba spin texture that is the origin of the switchable SREE \cite{Liebmann2016,Elmers2016,Rinaldi2018}. The SREE coefficient $\chi_s$ and dynamical magnetic susceptibility $\Sigma_{\rm B}$ are displayed as a function of the energy in Figs. \ref{fig3}(a) and (b), respectively. The OREE is also reported in Fig. \ref{fig3} and will be discussed in the next section. We notice that the Rashba-Edelstein effect for this material has only one independent component, where $\chi_{xy}=-\chi_{yx}=\chi_s$. The SREE coefficient reaches a local maximum around -0.3 eV below the gap, which is consistent with a recent work \cite{Leiva-Montecinos2025}. In contrast, the dynamical magnetic susceptibility in Fig. \ref{fig3}(b) exhibits a local maximum around -0.7 eV below the gap, demonstrating a similar behavior to the longitudinal conductivity within the same range [see Fig. \ref{fig2}(a)]. 

\begin{figure}[ht!]
	\includegraphics[width=\columnwidth]{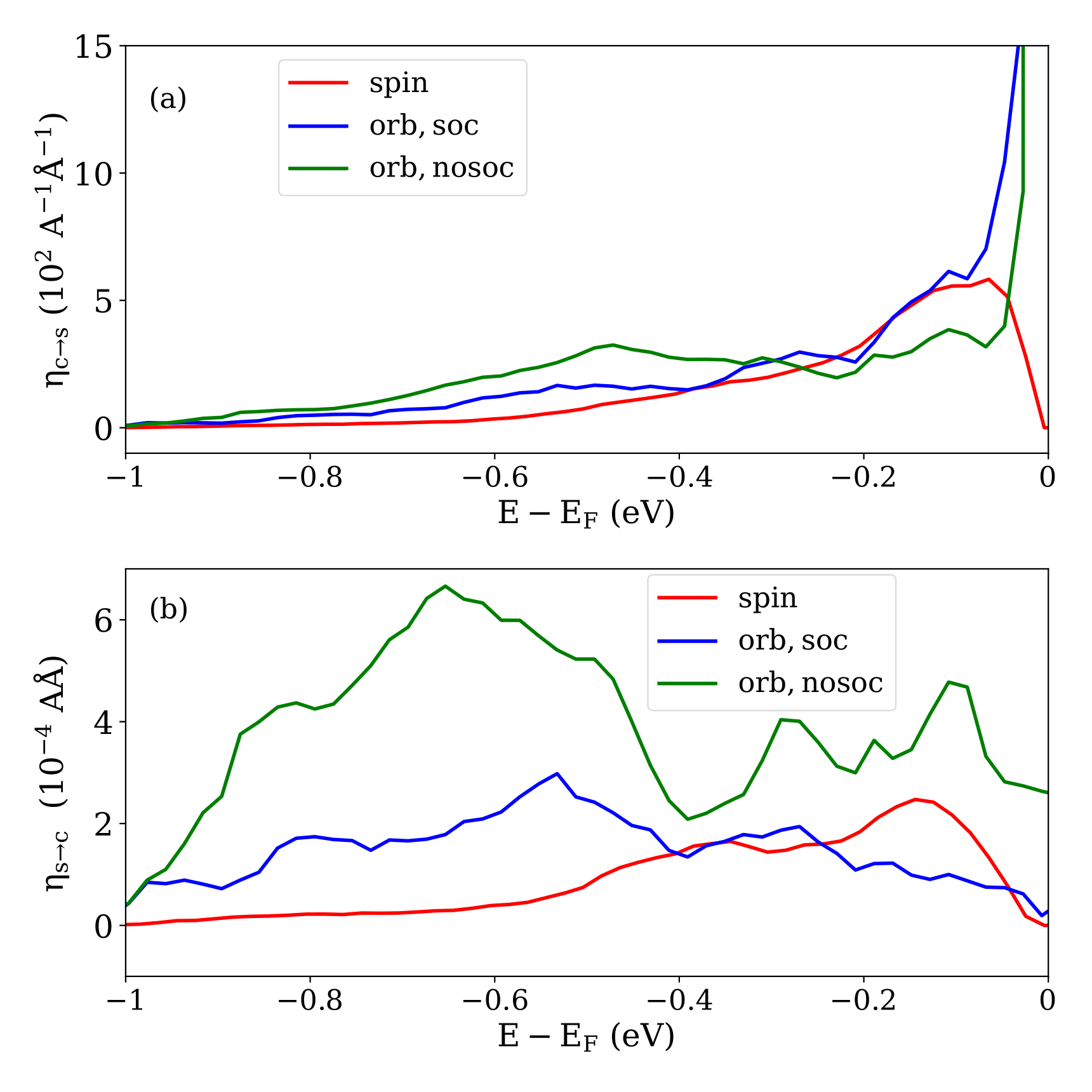}
    \caption{(Color online) (a) charge-to-spin (red) and charge-to-orbital conversion coefficients with (blue) and without spin-orbit coupling (green) as a function of the energy. (b) Spin-to-charge (red) and orbital-to-charge conversion coefficients with (blue) and without spin-orbit coupling (green) as a function of the energy.}
	\label{fig4}
	\end{figure}

Combining these different results, we obtain the charge-to-spin, $\eta_{\rm c\rightarrow s}$, and spin-to-charge conversion coefficients, $\eta_{\rm s\rightarrow c}$, reported in Fig. \ref{fig4}. These two coefficients exhibit a qualitatively similar behavior below the gap, first increasing and progressively decreasing away from the gap. The detailed structure of their energy dependence differs, though, since these terms are not Onsager-reciprocal, as discussed above. We emphasize that whereas the spin Hall angle scales as $\tau^{-1}$, the SREE coefficients are independent of $\tau$. As a consequence, $\theta_{\rm SH}$ is enhanced by the disorder in the material. 
 
\begin{figure}[ht!]
	\includegraphics[width=\columnwidth]{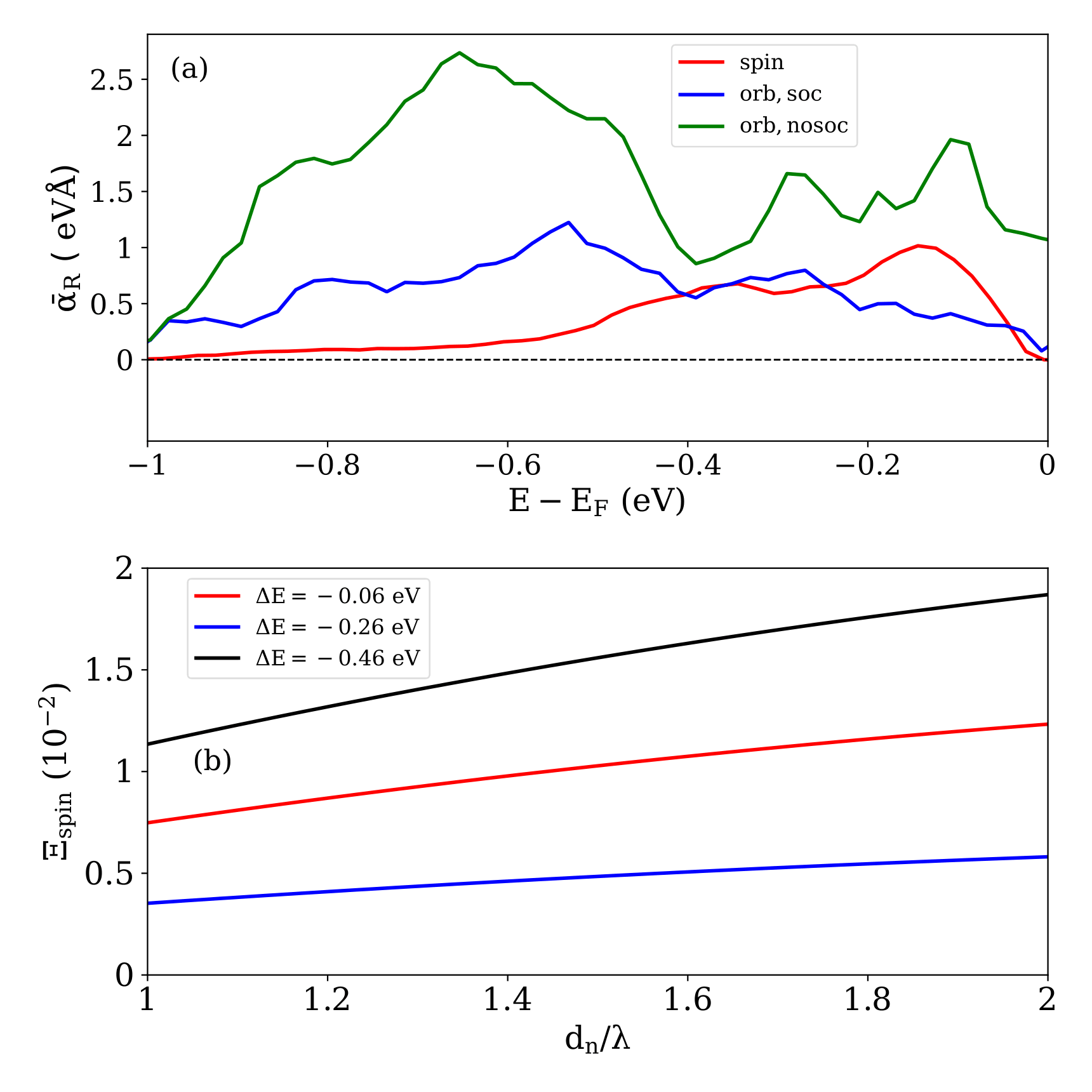}
    \caption{(Color online) (a) Effective spin Rashba parameter (red) estimated from the microscopic theory as a function of the energy. The orbital Rashba parameters with (blue) and without spin-orbit coupling (green) are also reported. (b) Ratio between the charge currents produced by SREE and OREE estimated from the drift diffusion model as a function of the ratio between the thickness of the ferroelectric conductor and the spin diffusion length $d_{\rm N}/\lambda$, for different values of the energy $\Delta \rm E=E-E_F$.}
	\label{fig5}
	\end{figure}

To estimate the relative magnitude between SHE and SREE in $\alpha$-GeTe, we need to estimate the \textit{effective} Rashba parameter $\bar{\alpha}_R$ entering Eq. \eqref{drift8}. This parameter is given by 
\begin{eqnarray}
\bar{\alpha}_R=\frac{\hbar}{e}\eta_{s\to c}=\mu_B\frac{\hbar}{e}\frac{\chi_s}{\Sigma_B},\label{alpha1}
\end{eqnarray}
and is reported on Fig. \ref{fig5}(a). Surprisingly, the effective Rashba parameter is two to three orders of magnitude smaller than what was reported in the literature \cite{DiSante2013}. We attribute this difference to cancellations between bands, away from the $\Gamma$ point, when we take into account all the intraband transitions at the Fermi level. Indeed, whereas the Rashba parameter defined in Eq. \eqref{eq:{SREE}} is valid close to the $\Gamma$ point, the realistic computation conducted in the present work accounts for {\em all} bands crossing a given Fermi level. This cancellation has also been verified previously, even at the low energy regime \cite{Leiva-Montecinos2025}. From this estimate, we compute the ratio $\Xi$ defined by Eq. \eqref{drift8} and illustrated in Fig. \ref{fig5}(b). Here we deduce a larger contribution from SREE rather than SHE, which remains relatively constant with the increase of the spin diffusion length and the magnitude of $\Delta \rm E$ (at least, in the $p$-doped regime). This behavior can be attributed to the large spin injection induced by the interplay between the current injected and the Rashba spin texture associated with the ferroelectric polarization. In the next section, we complement this analysis with the case of the orbital-to-charge conversion. 

\subsection{Orbital-to-charge interconversion}

To offer a comprehensive description of the transport of atomic angular momentum in $\alpha$-GeTe, we also compute the atomic orbital properties in Figs. \ref{fig2}, \ref{fig3} and \ref{fig4}. The OHE in Fig. \ref{fig2}(a) exhibits qualitatively the same energy dependence as that of SHE, although with a much larger magnitude, as often observed in metals with dominant d-orbital structure \cite{Salemi2022}. Spin-orbit coupling enhances the effect and, most importantly, no sign reversal is observed, contrary to SHE. A remarkable feature is the OHE plateau obtained in the gap of $\alpha$-GeTe. Such a plateau has been obtained in other materials, such as PbTe or MoS$_2$ \cite{Pezo2022}. In the latter case, it has been associated with higher-order topology \cite{Costa2023}. $\alpha$-GeTe is not a known topological insulator, and the study of its topological properties lies outside the scope of the present study. This plateau is nonetheless intriguing and certainly calls for further investigations. According to Fig. \ref{fig2}(b), the orbital Hall angle remains approximately constant as a function of the energy away from the gap. Besides, its magnitude is slightly larger than the spin Hall angle, especially in the absence of spin-orbit coupling. \par

As we can extract from Fig. \ref{fig3}(a), the energy dependence of the OREE is similar to that of SREE, but its overall magnitude is, again, generally larger, with a minor impact of the spin-orbit coupling. The orbital susceptibility in Fig. \ref{fig3}(b) remains smaller than the spin susceptibility, and it is enhanced when spin-orbit coupling is taken into account. Interestingly, the charge-to-orbital and orbital-to-charge interconversion coefficients in Fig. \ref{fig4} display completely different energy dependence, and the influence of spin-orbit coupling is strong on the latter, partially reducing the orbital-to-charge coefficient. As a result, the orbital Rashba parameter, which is sketched in Fig. \ref{fig5}(a), is also generally larger than the spin Rashba parameter and partially quenched by turning on the spin-orbit coupling.

\section{Discussion and Conclusion}\label{IV}

\begin{figure}[ht!]
	\includegraphics[width=\columnwidth]{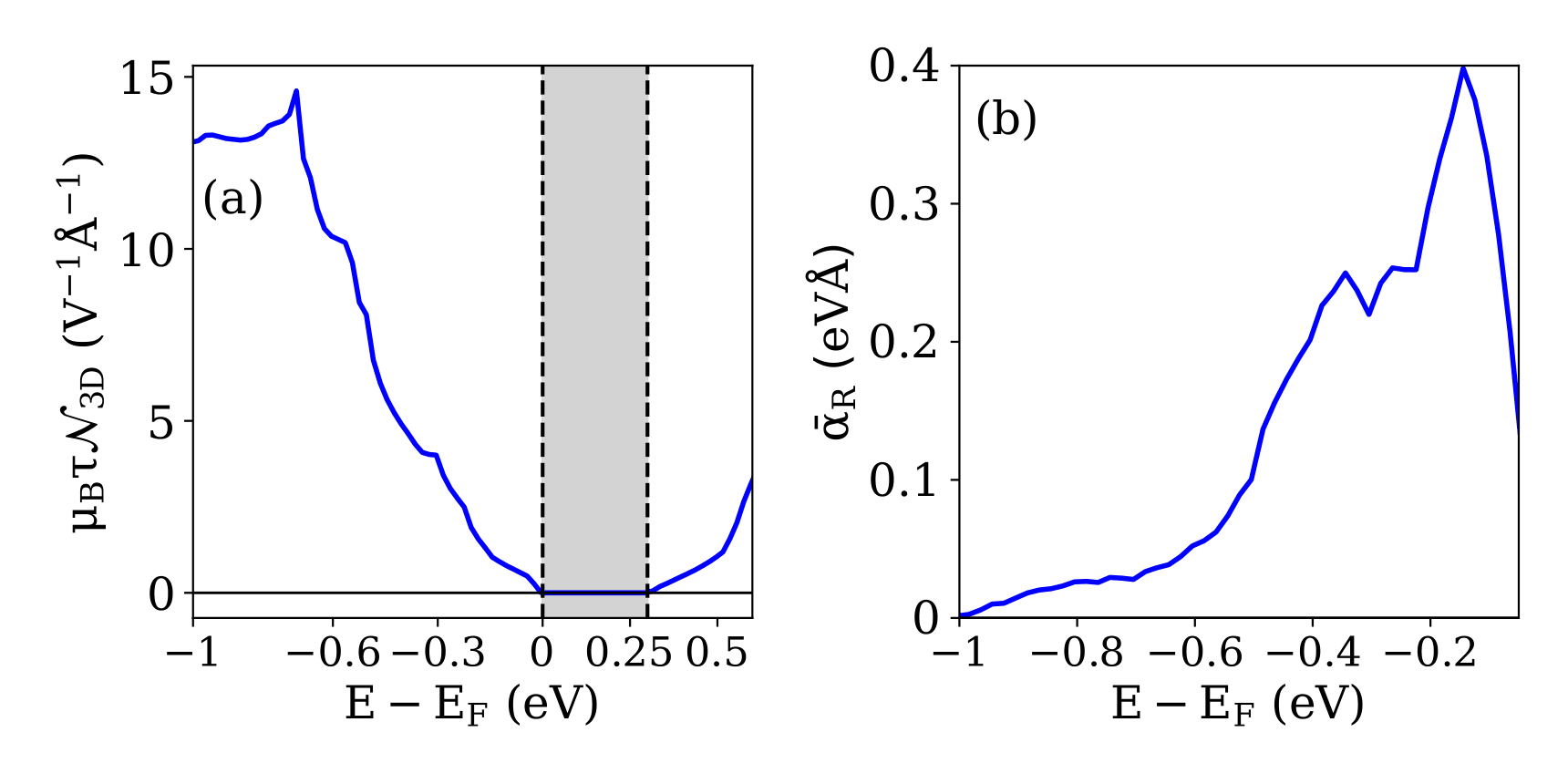}
    \caption{(Color online) (a) Density of states coefficient, modulated by the momentum scattering time as discussed in the main text. (b) Rashba parameter from this theory.}
	\label{fig6}
	\end{figure}

Let us now discuss the validity of our approach. In most previous literature, the direct and inverse SREE are connected through the inverse Rashba-Edelstein length \cite{Isshiki2020}
\begin{align}
    \lambda_{\rm IEE}&=\frac{\chi_s}{e\mathcal{N}},\label{dis1}
\end{align}

\noindent 
which satisfies $\lambda_{IEE}\simeq \bar{\alpha}_R\tau\hbar^{-1}$ \cite{Rojas-Sanchez2013b,Shen2014,Zhang2016n}. Then, the Rashba parameter from this model is given by

\begin{align}
    \bar{\alpha}_R&=\frac{\hbar\chi_s}{e\tau\mathcal{N}},\label{dis2}
\end{align}

\noindent
which is independent of $\tau$ as in our identity in Eq. \eqref{alpha1}. Notice that Eq. \eqref{dis2} comprises the density of states, whereas Eq. \eqref{alpha1} replaces it by the magnetic susceptibility. In view of both definitions and their explicit expressions in terms of $\Sigma_B$ and $\mathcal{N}$, we plot $\mu_B\tau\mathcal{N}$ as a function of the energy in Fig. \ref{fig6}(a). Compared to our theory in Fig. \ref{fig3}(b), these coefficients are about the same order of magnitude, and they share the overall same behavior in the energy range of the calculation, corroborating the robustness of our theory. To complement this evidence, we plot the Rashba parameter from the phenomenological model in Eq. \eqref{dis2}, Fig. \ref{fig6}(b), which yields the same qualitative agreement between this model and our theory.

In the general picture of this work and in the case of the SHE, it is worth mentioning that a large spin Hall conductivity does not guarantee a large $\theta_{\rm SH}$ (For instance, in the fundamental instance of Pt \cite{Guo2008}). Actually, there is a delicate compromise between the spin Hall and the longitudinal conductivity of the material. This connection is expected to lead to a large $\theta_{\rm SH}$ in the TaAs family of compounds, even though the spin Hall conductivity is orders of magnitude smaller than in Pt \cite{Sun2016}. We suggest analyzing the spin-to-charge conversion coefficients in this family arising from the SREE and comparing them to the SHE counterpart, complementing an earlier study that applied semiclassical considerations \cite{Johansson2018} and its measurement at room temperature in TaP \cite{Mendes2022}. As the tilting of the Weyl nodes and the number of Weyl point is in principle relevant to enhance the Rashba effect and the SHE, it is worth to consider as well the (re)evaluation of the spin-to-charge conversion in type II Weyl semimetals with noncentrosymmetric structure such as $\mathrm{WTe}_2$ \cite{Zhao2020} and $\mathrm{TaIrTe_4}$. In addition, it would also be interesting to follow the role of the surface and bulk states on the spin-to-charge conversion not only on topological semimetals, but also in topological insulators \cite{Jamali2015}. Furthermore, since the Rashba parameter can be tuned through strain \cite{Liu2021}, it is instructive to contrast the Rashba parameters coming from the minimal model formalism and the theory described in this article.

In conclusion, we establish a rigorous theoretical framework for defining spin-to-charge and orbital-to-charge interconversion coefficients in nonmagnetic materials lacking inversion symmetry. These coefficients are expressed in terms of macroscopic observables that are directly accessible experimentally. Applying this framework to the ferroelectric semiconductor $\alpha$-GeTe, we find that the effective Rashba parameter deviates from previous theoretical estimates, a discrepancy that we attribute to differences in the number of bands considered in the respective analyses. Incorporating the derived coefficients
into a standard drift–diffusion model reveals that the Rashba–Edelstein mechanism dominates over the spin and orbital Hall effects in $\alpha$-GeTe. This dominance arises primarily from the strong spin and orbital injections associated with the Rashba texture induced by ferroelectric
polarization, which exceed the corresponding spin and orbital Hall conductivities. Consequently, charge transport in $\alpha$-GeTe is governed predominantly by the ferroelectric polarization through the Rashba effect, rather than by bulk spin–orbit–coupling–induced Hall responses. It would be instructive to compare these results with those obtained for other ferroelectric systems, such as doped SnTe \cite{Chassot2024,Wang2020}, or HfO$_2$ \cite{Tao2017,Zhang2024}.

\begin{acknowledgments}
D.G.O. thanks Andrés Sa\'ul and Xiaobai Ning for insightful discussions. In addition, the authors also acknowledge Gen Tatara for valuable feedback on the theoretical side of this work. This work was supported by the grant AMX-22-RE-AB-043 of the Excellence Initiative of Aix-Marseille Université - A*Midex, a French "Investissements d'Avenir" program, by France 2030 government investment plan managed by the French National Research Agency under grants reference PEPR SPIN – [SPINTHEORY] ANR-22-EXSP-0009 and [OXIMOR] ANR-24-EXSP-0011, and by the EIC Pathfinder OPEN grant 101129641 “OBELIX”.
\end{acknowledgments}

\bibliography{Biblio_scc} 

@article{Yue2018,
  title = {Spin-to-Charge Conversion in Bi Films and $\mathrm{Bi}/\mathrm{Ag}$ Bilayers},
  author = {Yue, Di and Lin, Weiwei and Li, Jiajia and Jin, Xiaofeng and Chien, C. L.},
  journal = {Phys. Rev. Lett.},
  volume = {121},
  issue = {3},
  pages = {037201},
  numpages = {6},
  year = {2018},
  month = {Jul},
  publisher = {American Physical Society},
  doi = {10.1103/PhysRevLett.121.037201},
  url = {https://link.aps.org/doi/10.1103/PhysRevLett.121.037201}
}

@article{Bonbien2020,
   author = {V. Bonbien and A. Manchon},
   doi = {10.1103/PhysRevB.102.085113},
   issn = {24699969},
   journal = {Physical Review B},
   pages = {085113},
   title = {Symmetrized decomposition of the Kubo-Bastin formula},
   volume = {102},
   year = {2020}
}

@article{Shi2025,
   author = {Shuyuan Shi and Enlong Liu and Fanrui Hu and Guoyi Shi and Aurélien Manchon and Hyunsoo Yang},
   doi = {10.1103/PhysRevB.111.094433},
   issn = {2469-9950},
   issue = {9},
   journal = {Physical Review B},
   month = {3},
   pages = {094433},
   title = {Nonreciprocal spin-charge interconversion in topological insulator/ferromagnet heterostructures},
   volume = {111},
   url = {https://link.aps.org/doi/10.1103/PhysRevB.111.094433},
   year = {2025}
}

@article{Chen2015c,
   author = {Kai Chen and Shufeng Zhang},
   doi = {10.1103/PhysRevLett.114.126602},
   issn = {10797114},
   journal = {Physical Review Letters},
   pages = {126602},
   pmid = {25860763},
   title = {Spin pumping in the presence of spin-orbit coupling},
   volume = {114},
   year = {2015}
}

@article{Cheng2022,
  title = {Coherent Picture on the Pure Spin Transport between $\mathrm{Ag}/\mathrm{Bi}$ and Ferromagnets},
  author = {Cheng, J. and Miao, B. F. and Liu, Z. and Yang, M. and He, K. and Zeng, Y. L. and Niu, H. and Yang, X. and Wang, Z. Q. and Hong, X. H. and Fu, S. J. and Sun, L. and Liu, Y. and Wu, Y. Z. and Yuan, Z. and Ding, H. F.},
  journal = {Phys. Rev. Lett.},
  volume = {129},
  issue = {9},
  pages = {097203},
  numpages = {6},
  year = {2022},
  month = {Aug},
  publisher = {American Physical Society},
  doi = {10.1103/PhysRevLett.129.097203},
  url = {https://link.aps.org/doi/10.1103/PhysRevLett.129.097203}
}

@article{Yu2020,
  title = {Fingerprint of the inverse Rashba-Edelstein effect at heavy-metal/Cu interfaces},
  author = {Yu, Rui and Miao, Bingfeng and Liu, Qi and He, Kang and Xue, Weishan and Sun, Liang and Wu, Mingzhong and Wu, Yizheng and Yuan, Zhe and Ding, Haifeng},
  journal = {Phys. Rev. B},
  volume = {102},
  issue = {14},
  pages = {144415},
  numpages = {7},
  year = {2020},
  month = {Oct},
  publisher = {American Physical Society},
  doi = {10.1103/PhysRevB.102.144415},
  url = {https://link.aps.org/doi/10.1103/PhysRevB.102.144415}
}

@article{Zhao2020,
  title = {Observation of charge to spin conversion in Weyl semimetal ${\mathrm{WTe}}_{2}$ at room temperature},
  author = {Zhao, Bing and Khokhriakov, Dmitrii and Zhang, Yang and Fu, Huixia and Karpiak, Bogdan and Hoque, Anamul Md. and Xu, Xiaoguang and Jiang, Yong and Yan, Binghai and Dash, Saroj P.},
  journal = {Phys. Rev. Res.},
  volume = {2},
  issue = {1},
  pages = {013286},
  numpages = {8},
  year = {2020},
  month = {Mar},
  publisher = {American Physical Society},
  doi = {10.1103/PhysRevResearch.2.013286},
  url = {https://link.aps.org/doi/10.1103/PhysRevResearch.2.013286}
}

@article{Vesta2011,
author = "Momma, Koichi and Izumi, Fujio",
title = "{{\it VESTA3} for three-dimensional visualization of crystal, volumetric and morphology data}",
journal = "Journal of Applied Crystallography",
year = "2011",
volume = "44",
number = "6",
pages = "1272--1276",
month = "Dec",
doi = {10.1107/S0021889811038970},
url = {https://doi.org/10.1107/S0021889811038970},
}

@article{Elmers2016,
   author = {H. J. Elmers and R. Wallauer and M. Liebmann and J. Kellner and M. Morgenstern and R. N. Wang and J. E. Boschker and R. Calarco and J. Sánchez-Barriga and O. Rader and D. Kutnyakhov and S. V. Chernov and K. Medjanik and C. Tusche and M. Ellguth and H. Volfova and St Borek and J. Braun and J. Minár and H. Ebert and G. Schönhense},
   doi = {10.1103/PhysRevB.94.201403},
   issn = {24699969},
   journal = {Physical Review B},
   pages = {201403(R)},
   title = {Spin mapping of surface and bulk Rashba states in ferroelectric $\alpha$-GeTe(111) films},
   volume = {94},
   year = {2016}
}

@article{Rinaldi2018,
   author = {Christian Rinaldi and Sara Varotto and Marco Asa and Jagoda Sławińska and Jun Fujii and Giovanni Vinai and Stefano Cecchi and Domenico Di Sante and Raffaella Calarco and Ivana Vobornik and Giancarlo Panaccione and Silvia Picozzi and Riccardo Bertacco},
   doi = {10.1021/acs.nanolett.7b04829},
   issn = {15306992},
   journal = {Nano Letters},
   pages = {2751},
   pmid = {29380606},
   title = {Ferroelectric Control of the Spin Texture in GeTe},
   volume = {18},
   year = {2018}
}

@article{Isshiki2020,
  title = {Phenomenological model for the direct and inverse Edelstein effects},
  author = {Isshiki, Hironari and Muduli, Prasanta and Kim, Junyeon and Kondou, Kouta and Otani, YoshiChika},
  journal = {Phys. Rev. B},
  volume = {102},
  issue = {18},
  pages = {184411},
  numpages = {5},
  year = {2020},
  month = {Nov},
  publisher = {American Physical Society},
  doi = {10.1103/PhysRevB.102.184411},
  url = {https://link.aps.org/doi/10.1103/PhysRevB.102.184411}}

@article{Jo2018,
   author = {Daegeun Jo and Dongwook Go and Hyun-woo Lee},
   doi = {10.1103/PhysRevB.98.214405},
   journal = {Physical Review B},
   pages = {214405},
   publisher = {American Physical Society},
   title = {Gigantic intrinsic orbital Hall effects in weakly spin-orbit coupled metals},
   volume = {98},
   year = {2018}
}

@article{Go2025,
   author = {Dongwook Go and Kazuya Ando and Armando Pezo and Stefan Blügel and Aurélien Manchon and Yuriy Mokrousov},
   doi = {10.1103/PhysRevB.111.L140409},
   issn = {24699969},
   issue = {14},
   journal = {Physical Review B},
   month = {4},
   pages = {L140409},
   publisher = {American Physical Society},
   title = {Orbital pumping by magnetization dynamics in ferromagnets},
   volume = {111},
   year = {2025}
}

@article{Leiva-Montecinos2025,
   author = {Sergio Leiva-Montecinos and Libor Vojáček and Jing Li and Mairbek Chshiev and Laurent Vila and Ingrid Mertig and Annika Johansson},
   journal = {arXiv:2505.21340},
   month = {8},
   title = {Current-induced spin and orbital polarization in the ferroelectric Rashba semiconductor GeTe},
   url = {http://arxiv.org/abs/2505.21340},
   year = {2025}
}

@article{Salemi2022,
   author = {Leandro Salemi and Peter M. Oppeneer},
   doi = {10.1103/physrevmaterials.6.095001},
   journal = {Physical Review Materials},
   pages = {095001},
   publisher = {American Physical Society},
   title = {First-principles theory of intrinsic spin and orbital Hall and Nernst effects in metallic monoatomic crystals},
   volume = {6},
   year = {2022}
}

@article{Zhang2016n,
   author = {S. Zhang and A. Fert},
   doi = {10.1103/PhysRevB.94.184423},
   issn = {24699969},
   journal = {Physical Review B},
   pages = {184423},
   title = {Conversion between spin and charge currents with topological insulators},
   volume = {94},
   year = {2016}
}

@article{Rojas-Sanchez2013b,
   author = {J C Rojas-Sánchez and L Vila and G Desfonds and S Gambarelli and J.-P. Attane and J M De Teresa and C Magén and A. Fert},
   doi = {10.1038/ncomms3944},
   issn = {2041-1723},
   journal = {Nature Communications},
   month = {1},
   pages = {2944},
   pmid = {24343336},
   title = {Spin-to-charge conversion using Rashba coupling at the interface between non-magnetic materials.},
   volume = {4},
   url = {http://www.ncbi.nlm.nih.gov/pubmed/24343336},
   year = {2013}
}

@article{Chassot2024,
   author = {Frédéric Chassot and Aki Pulkkinen and Geoffroy Kremer and Tetiana Zakusylo and Gauthier Krizman and Mahdi Hajlaoui and J. Hugo Dil and Juraj Krempaský and Ján Minár and Gunther Springholz and Claude Monney},
   doi = {10.1021/acs.nanolett.3c03280},
   issn = {15306992},
   issue = {1},
   journal = {Nano Letters},
   month = {1},
   pages = {82-88},
   pmid = {38109843},
   publisher = {American Chemical Society},
   title = {Persistence of Structural Distortion and Bulk Band Rashba Splitting in SnTe above Its Ferroelectric Critical Temperature},
   volume = {24},
   year = {2024}
}

@article{Costa2023,
   author = {Marcio Costa and Bruno Focassio and Luis M. Canonico and Tarik P. Cysne and Gabriel R. Schleder and R. B. Muniz and Adalberto Fazzio and Tatiana G. Rappoport},
   doi = {10.1103/PhysRevLett.130.116204},
   issn = {10797114},
   issue = {11},
   journal = {Physical Review Letters},
   month = {3},
   pages = {116204},
   pmid = {37001112},
   publisher = {American Physical Society},
   title = {Connecting Higher-Order Topology with the Orbital Hall Effect in Monolayers of Transition Metal Dichalcogenides},
   volume = {130},
   year = {2023}
}

@article{Pezo2025,
   author = {Armando Pezo and Dongwook Go and Yuriy Mokrousov and Henri Jaffrès and Aurélien Manchon},
   doi = {10.1103/PhysRevB.111.134424},
   issn = {24699969},
   issue = {13},
   journal = {Physical Review B},
   month = {4},
   pages = {134424},
   publisher = {American Physical Society},
   title = {Adiabatic spin and orbital pumping in metallic heterostructures},
   volume = {111},
   year = {2025}
}

@article{Han2025,
   author = {Seungyun Han and Hye Won Ko and Jung Hyun Oh and Hyun Woo Lee and Kyung Jin Lee and Kyoung Whan Kim},
   doi = {10.1103/PhysRevLett.134.036305},
   issn = {10797114},
   issue = {3},
   journal = {Physical Review Letters},
   month = {1},
   pages = {036305},
   pmid = {39927955},
   publisher = {American Physical Society},
   title = {Orbital Pumping Incorporating Both Orbital Angular Momentum and Position},
   volume = {134},
   year = {2025}
}

@article{Pezo2022,
   author = {Armando Pezo and Diego Garcia Ovalle and Aurelien Manchon},
   doi = {10.1103/PhysRevB.106.104414},
   issn = {24699969},
   journal = {Physical Review B},
   pages = {104414},
   publisher = {American Physical Society},
   title = {Orbital Hall effect in crystals: inter-atomic versus intra-atomic contributions},
   volume = {106},
   url = {http://arxiv.org/abs/2201.05807},
   year = {2022}
}

@article{Go2017,
   author = {Dongwook Go and Jan-philipp Hanke and Patrick M Buhl and Frank Freimuth and Gustav Bihlmayer and Hyun-woo Lee and Yuriy Mokrousov and Stefan Blügel},
   doi = {10.1038/srep46742},
   journal = {Scientific Reports},
   pages = {46742},
   publisher = {Nature Publishing Group},
   title = {Toward surface orbitronics : giant orbital magnetism from the orbital Rashba effect at the surface of sp-metals},
   volume = {7},
   url = {http://dx.doi.org/10.1038/srep46742},
   year = {2017}
}

@article{Choi2023,
   author = {Young Gwan Choi and Daegeun Jo and Kyung Hun Ko and Dongwook Go and Kyung Han Kim and Hee Gyum Park and Changyoung Kim and Byoung Chul Min and Gyung Min Choi and Hyun Woo Lee},
   doi = {10.1038/s41586-023-06101-9},
   issn = {14764687},
   issue = {7968},
   journal = {Nature},
   month = {7},
   pages = {52-56},
   pmid = {37407680},
   publisher = {Nature Research},
   title = {Observation of the orbital Hall effect in a light metal Ti},
   volume = {619},
   year = {2023}
}

@article{Go2018,
   author = {Dongwook Go and Daegeun Jo and Changyoung Kim and Hyun Woo Lee},
   doi = {10.1103/PhysRevLett.121.086602},
   isbn = {6082744330},
   issn = {10797114},
   journal = {Physical Review Letters},
   pages = {086602},
   publisher = {American Physical Society},
   title = {Intrinsic Spin and Orbital Hall Effects from Orbital Texture},
   volume = {121},
   url = {https://doi.org/10.1103/PhysRevLett.121.086602},
   year = {2018}
}

@article{Bihlmayer2022,
   author = {Gustav Bihlmayer and Paul Noël and Denis V. Vyalikh and Evgueni V. Chulkov and Aurélien Manchon},
   doi = {10.1038/s42254-022-00490-y},
   isbn = {4225402200490},
   journal = {Nature Reviews Physics},
   pages = {642},
   title = {Rashba-like physics in condensed matter},
   volume = {4},
   year = {2022}
}

@article{Dyakonov1971,
   author = {M I D'yakonov and V I Perel'},
   doi = {10.1063/1.1614421},
   isbn = {00036951},
   issn = {0021-3640},
   journal = {JETP Letters},
   pages = {467},
   title = {Possibility of orienting electron spins with current},
   url = {http://www.jetpletters.ac.ru/ps/1587/article_24366.shtml},
   year = {1971},
}

@article{Lesne2016,
author={Lesne, E.
and Fu, Yu
and Oyarzun, S.
and Rojas-S{\'a}nchez, J. C.
and Vaz, D. C.
and Naganuma, H.
and Sicoli, G.
and Attan{\'e}, J.-P.
and Jamet, M.
and Jacquet, E.
and George, J.-M.
and Barth{\'e}l{\'e}my, A.
and Jaffr{\`e}s, H.
and Fert, A.
and Bibes, M.
and Vila, L.},
title={Highly efficient and tunable spin-to-charge conversion through Rashba coupling at oxide interfaces},
journal={Nature Materials},
year={2016},
month={Dec},
day={01},
volume={15},
number={12},
pages={1261-1266},
issn={1476-4660},
doi={10.1038/nmat4726},
url={https://doi.org/10.1038/nmat4726}
}

@article{Mendes2022,
author = {Mendes, Joaquim B. S. and Vieira, Andriele S. and Cunha, Rafael O. and Ferreira, Sukarno O. and dos Reis, Ricardo D. and Schmidt, Marcus and Nicklas, Michael and Rezende, Sergio M. and Azevedo, Antonio},
title = {Efficient Spin-to-Charge Interconversion in Weyl Semimetal TaP at Room Temperature},
journal = {Advanced Materials Interfaces},
volume = {9},
number = {36},
pages = {2201716},
doi = {https://doi.org/10.1002/admi.202201716},
url = {https://onlinelibrary.wiley.com/doi/abs/10.1002/admi.202201716},
year = {2022}
}

@article{Vaz2019,
author={Vaz, Diogo C.
and No{\"e}l, Paul
and Johansson, Annika
and G{\"o}bel, B{\"o}rge
and Bruno, Flavio Y.
and Singh, Gyanendra
and McKeown-Walker, Siobhan
and Trier, Felix
and Vicente-Arche, Luis M.
and Sander, Anke
and Valencia, Sergio
and Bruneel, Pierre
and Vivek, Manali
and Gabay, Marc
and Bergeal, Nicolas
and Baumberger, Felix
and Okuno, Hanako
and Barth{\'e}l{\'e}my, Agn{\`e}s
and Fert, Albert
and Vila, Laurent
and Mertig, Ingrid
and Attan{\'e}, Jean-Philippe
and Bibes, Manuel},
title={Mapping spin--charge conversion to the band structure in a topological oxide two-dimensional electron gas},
journal={Nature Materials},
year={2019},
month={Nov},
day={01},
volume={18},
number={11},
pages={1187-1193},
issn={1476-4660},
doi={10.1038/s41563-019-0467-4},
url={https://doi.org/10.1038/s41563-019-0467-4}
}

@article{Picozzi2014,
   author = {Silvia Picozzi},
   doi = {10.3389/fphy.2014.00010},
   issn = {2296424X},
   journal = {Frontiers in Physics},
   pages = {1},
   title = {Ferroelectric Rashba semiconductors as a novel class of multifunctional materials},
   volume = {2},
   year = {2014},
}

@article{Liebmann2016,
   author = {Marcus Liebmann and Christian Rinaldi and Domenico Di Sante and Jens Kellner and Christian Pauly and Rui Ning Wang and Jos Emiel Boschker and Alessandro Giussani and Stefano Bertoli and Matteo Cantoni and Lorenzo Baldrati and Marco Asa and Ivana Vobornik and Giancarlo Panaccione and Dmitry Marchenko and Jaime Sánchez-Barriga and Oliver Rader and Raffaella Calarco and Silvia Picozzi and Riccardo Bertacco and Markus Morgenstern},
   doi = {10.1002/adma.201503459},
   issn = {15214095},
   journal = {Advanced Materials},
   pages = {560},
   title = {Giant Rashba-Type Spin Splitting in Ferroelectric GeTe(111)},
   volume = {28},
   year = {2016},
}

@article{Shchelushkin2005b,
   author = {R. Shchelushkin and Arne Brataas},
   doi = {10.1103/PhysRevB.71.045123},
   issn = {1098-0121},
   journal = {Physical Review B},
   month = {1},
   pages = {045123},
   title = {Spin Hall effects in diffusive normal metals},
   volume = {71},
   url = {http://link.aps.org/doi/10.1103/PhysRevB.71.045123},
   year = {2005},
}

@article{Brataas2002,
   author = {Arne Brataas and Yaroslav Tserkovnyak and G. E. W. Bauer and Bertrand Halperin},
   doi = {10.1103/PhysRevB.66.060404},
   issn = {0163-1829},
   journal = {Physical Review B},
   month = {8},
   pages = {060404},
   title = {Spin battery operated by ferromagnetic resonance},
   volume = {66},
   url = {http://link.aps.org/doi/10.1103/PhysRevB.66.060404},
   year = {2002},
}

@article{Tserkovnyak2002b,
   author = {Yaroslav Tserkovnyak and Arne Brataas and G. E. W. Bauer},
   doi = {10.1103/PhysRevB.66.224403},
   issn = {0163-1829},
   journal = {Physical Review B},
   month = {12},
   pages = {224403},
   title = {Spin pumping and magnetization dynamics in metallic multilayers},
   volume = {66},
   url = {http://link.aps.org/doi/10.1103/PhysRevB.66.224403},
   year = {2002},
}

@article{Sharma2019,
author = {Pankaj Sharma  and Fei-Xiang Xiang  and Ding-Fu Shao  and Dawei Zhang  and Evgeny Y. Tsymbal  and Alex R. Hamilton  and Jan Seidel },
title = {A room-temperature ferroelectric semimetal},
journal = {Science Advances},
volume = {5},
number = {7},
pages = {eaax5080},
year = {2019},
doi = {10.1126/sciadv.aax5080},
URL = {https://www.science.org/doi/abs/10.1126/sciadv.aax5080},
eprint = {https://www.science.org/doi/pdf/10.1126/sciadv.aax5080}}

@article{Liu2021,
  title = {Tuning Rashba effect, band inversion, and spin-charge conversion of Janus $X{\mathrm{Sn}}_{2}Y$ monolayers via an external field},
  author = {Liu, Ming-Yang and Gong, Long and He, Yao and Cao, Chao},
  journal = {Phys. Rev. B},
  volume = {103},
  issue = {7},
  pages = {075421},
  numpages = {17},
  year = {2021},
  month = {Feb},
  publisher = {American Physical Society},
  doi = {10.1103/PhysRevB.103.075421},
  url = {https://link.aps.org/doi/10.1103/PhysRevB.103.075421}
}

@article{Ishizaka2011,
   author = {K Ishizaka and M S Bahramy and H Murakawa and M Sakano and T Shimojima and T Sonobe and K Koizumi and S Shin and H Miyahara and A Kimura and K Miyamoto and T Okuda and H Namatame and M Taniguchi and Ryotaro Arita and N Nagaosa and K Kobayashi and Y Murakami and R Kumai and Y Kaneko and Y Onose and Y Tokura},
   issn = {1476-1122},
   journal = {Nature materials},
   month = {7},
   note = {10.1038/nmat3051},
   pages = {521},
   publisher = {Nature Publishing Group},
   title = {Giant Rashba-type spin splitting in bulk BiTeI},
   volume = {10},
   url = {http://dx.doi.org/10.1038/nmat3051},
   year = {2011},
}

@article{RojasSanchez2013b,
   author = {J C Rojas-Sánchez and L Vila and G Desfonds and S Gambarelli and J.-P. Attane and J M De Teresa and C Magén and A. Fert},
   doi = {10.1038/ncomms3944},
   issn = {2041-1723},
   journal = {Nature Communications},
   month = {1},
   pages = {2944},
   pmid = {24343336},
   title = {Spin-to-charge conversion using Rashba coupling at the interface between non-magnetic materials.},
   volume = {4},
   url = {http://www.ncbi.nlm.nih.gov/pubmed/24343336},
   year = {2013},
}

@article{RojasSanchez2016,
  title = {Spin to Charge Conversion at Room Temperature by Spin Pumping into a New Type of Topological Insulator: $\ensuremath{\alpha}$-Sn Films},
  author = {Rojas-S\'anchez, J.-C. and Oyarz\'un, S. and Fu, Y. and Marty, A. and Vergnaud, C. and Gambarelli, S. and Vila, L. and Jamet, M. and Ohtsubo, Y. and Taleb-Ibrahimi, A. and Le F\`evre, P. and Bertran, F. and Reyren, N. and George, J.-M. and Fert, A.},
  journal = {Phys. Rev. Lett.},
  volume = {116},
  issue = {9},
  pages = {096602},
  numpages = {6},
  year = {2016},
  month = {Mar},
  publisher = {American Physical Society},
  doi = {10.1103/PhysRevLett.116.096602},
  url = {https://link.aps.org/doi/10.1103/PhysRevLett.116.096602}
}

@article{Wang2017b,
author={Wang, Yi
and Ramaswamy, Rajagopalan
and Motapothula, Mallikarjuna
and Narayanapillai, Kulothungasagaran
and Zhu, Dapeng
and Yu, Jiawei
and Venkatesan, Thirumalai
and Yang, Hyunsoo},
title={Room-Temperature Giant Charge-to-Spin Conversion at the SrTiO3--LaAlO3 Oxide Interface},
journal={Nano Letters},
year={2017},
month={Dec},
day={13},
publisher={American Chemical Society},
volume={17},
number={12},
pages={7659-7664},
issn={1530-6984},
doi={10.1021/acs.nanolett.7b03714},
url={https://doi.org/10.1021/acs.nanolett.7b03714}
}

@article{Varotto2022,
author={Varotto, Sara
and Johansson, Annika
and G{\"o}bel, B{\"o}rge
and Vicente-Arche, Luis M.
and Mallik, Srijani
and Br{\'e}hin, Julien
and Salazar, Rapha{\"e}l
and Bertran, Fran{\c{c}}ois
and F{\`e}vre, Patrick Le
and Bergeal, Nicolas
and Rault, Julien
and Mertig, Ingrid
and Bibes, Manuel},
title={Direct visualization of Rashba-split bands and spin/orbital-charge interconversion at KTaO3 interfaces},
journal={Nature Communications},
year={2022},
month={Oct},
day={18},
volume={13},
number={1},
pages={6165},
issn={2041-1723},
doi={10.1038/s41467-022-33621-1},
url={https://doi.org/10.1038/s41467-022-33621-1}
}

@article{Calavalle2022,
author={Calavalle, Francesco
and Su{\'a}rez-Rodr{\'i}guez, Manuel
and Mart{\'i}n-Garc{\'i}a, Beatriz
and Johansson, Annika
and Vaz, Diogo C.
and Yang, Haozhe
and Maznichenko, Igor V.
and Ostanin, Sergey
and Mateo-Alonso, Aurelio
and Chuvilin, Andrey
and Mertig, Ingrid
and Gobbi, Marco
and Casanova, F{\`e}lix
and Hueso, Luis E.},
title={Gate-tuneable and chirality-dependent charge-to-spin conversion in tellurium nanowires},
journal={Nature Materials},
year={2022},
month={May},
day={01},
volume={21},
number={5},
pages={526-532},
issn={1476-4660},
doi={10.1038/s41563-022-01211-7},
url={https://doi.org/10.1038/s41563-022-01211-7}
}

@article{Noel2020,
   author = {Paul Noël and Felix Trier and Luis M Vicente Arche and Julien Bréhin and Diogo C Vaz and Vincent Garcia and Stéphane Fusil and Agnès Barthélémy and Laurent Vila and Manuel Bibes and Jean-philippe Attané},
   doi = {10.1038/s41586-020-2197-9},
   issn = {1476-4687},
   journal = {Nature},
   pages = {483},
   publisher = {Springer US},
   title = {Non-volatile electric control of spin – charge conversion in a SrTiO 3 Rashba system},
   volume = {580},
   url = {http://dx.doi.org/10.1038/s41586-020-2197-9},
   year = {2020}
}

@article{Arche2022,
author = {Vicente-Arche, Luis M. and Bréhin, Julien and Varotto, Sara and Cosset-Cheneau, Maxen and Mallik, Srijani and Salazar, Raphaël and Noël, Paul and Vaz, Diogo C. and Trier, Felix and Bhattacharya, Suvam and Sander, Anke and Le Fèvre, Patrick and Bertran, François and Saiz, Guilhem and Ménard, Gerbold and Bergeal, Nicolas and Barthélémy, Agnès and Li, Hai and Lin, Chia-Ching and Nikonov, Dmitri E. and Young, Ian A. and Rault, Julien E. and Vila, Laurent and Attané, Jean-Philippe and Bibes, Manuel},
title = {Spin–Charge Interconversion in KTaO3 2D Electron Gases},
journal = {Advanced Materials},
volume = {33},
number = {43},
pages = {2102102},
doi = {https://doi.org/10.1002/adma.202102102},
url = {https://onlinelibrary.wiley.com/doi/abs/10.1002/adma.202102102},
year = {2021}
}

@article{Jafari2022,
  title = {Ferroelectric control of charge-to-spin conversion in ${\mathrm{WTe}}_{2}$},
  author = {Jafari, Homayoun and Roy, Arunesh and S\l{}awi\ifmmode \acute{n}\else \'{n}\fi{}ska, Jagoda},
  journal = {Phys. Rev. Mater.},
  volume = {6},
  issue = {9},
  pages = {L091404},
  numpages = {6},
  year = {2022},
  month = {Sep},
  publisher = {American Physical Society},
  doi = {10.1103/PhysRevMaterials.6.L091404},
  url = {https://link.aps.org/doi/10.1103/PhysRevMaterials.6.L091404}
}

@article{Jiao2013,
  title = {Spin Backflow and ac Voltage Generation by Spin Pumping and the Inverse Spin Hall Effect},
  author = {Jiao, HuJun and Bauer, Gerrit E. W.},
  journal = {Phys. Rev. Lett.},
  volume = {110},
  issue = {21},
  pages = {217602},
  numpages = {5},
  year = {2013},
  month = {May},
  publisher = {American Physical Society},
  doi = {10.1103/PhysRevLett.110.217602},
  url = {https://link.aps.org/doi/10.1103/PhysRevLett.110.217602}
}

@article{RojasSanchez2014,
  title = {Spin Pumping and Inverse Spin Hall Effect in Platinum: The Essential Role of Spin-Memory Loss at Metallic Interfaces},
  author = {Rojas-S\'anchez, J.-C. and Reyren, N. and Laczkowski, P. and Savero, W. and Attan\'e, J.-P. and Deranlot, C. and Jamet, M. and George, J.-M. and Vila, L. and Jaffr\`es, H.},
  journal = {Phys. Rev. Lett.},
  volume = {112},
  issue = {10},
  pages = {106602},
  numpages = {5},
  year = {2014},
  month = {Mar},
  publisher = {American Physical Society},
  doi = {10.1103/PhysRevLett.112.106602},
  url = {https://link.aps.org/doi/10.1103/PhysRevLett.112.106602}
}

@article{Mosendz2010b,
  title = {Quantifying Spin Hall Angles from Spin Pumping: Experiments and Theory},
  author = {Mosendz, O. and Pearson, J. E. and Fradin, F. Y. and Bauer, G. E. W. and Bader, S. D. and Hoffmann, A.},
  journal = {Phys. Rev. Lett.},
  volume = {104},
  issue = {4},
  pages = {046601},
  numpages = {4},
  year = {2010},
  month = {Jan},
  publisher = {American Physical Society},
  doi = {10.1103/PhysRevLett.104.046601},
  url = {https://link.aps.org/doi/10.1103/PhysRevLett.104.046601}
}

@article{Mosendz2010,
  title = {Detection and quantification of inverse spin Hall effect from spin pumping in permalloy/normal metal bilayers},
  author = {Mosendz, O. and Vlaminck, V. and Pearson, J. E. and Fradin, F. Y. and Bauer, G. E. W. and Bader, S. D. and Hoffmann, A.},
  journal = {Phys. Rev. B},
  volume = {82},
  issue = {21},
  pages = {214403},
  numpages = {10},
  year = {2010},
  month = {Dec},
  publisher = {American Physical Society},
  doi = {10.1103/PhysRevB.82.214403},
  url = {https://link.aps.org/doi/10.1103/PhysRevB.82.214403}
}

@article{Jamali2015,
   author = {Mahdi Jamali and Joon Sue Lee and Jong Seok Jeong and Farzad Mahfouzi and Yang Lv and Zhengyang Zhao and Branislav Nikolic and K. Andre Mkhoyan and Nitin Samarth and Jian-Ping Wang},
   doi = {10.1021/acs.nanolett.5b03274},
   issn = {1530-6984},
   journal = {Nano Letters},
   pages = {7126},
   title = {Giant Spin Pumping and Inverse Spin Hall Effect in the Presence of Surface Spin-Orbit Coupling of Topological Insulator Bi2Se3},
   volume = {15},
   url = {http://pubsdc3.acs.org/doi/10.1021/acs.nanolett.5b03274},
   year = {2015},
}

@article{Johansson2018,
  title = {Edelstein effect in Weyl semimetals},
  author = {Johansson, Annika and Henk, J\"urgen and Mertig, Ingrid},
  journal = {Phys. Rev. B},
  volume = {97},
  issue = {8},
  pages = {085417},
  numpages = {8},
  year = {2018},
  month = {Feb},
  publisher = {American Physical Society},
  doi = {10.1103/PhysRevB.97.085417},
  url = {https://link.aps.org/doi/10.1103/PhysRevB.97.085417}
}

@article{Sun2016,
  title = {Strong Intrinsic Spin Hall Effect in the TaAs Family of Weyl Semimetals},
  author = {Sun, Yan and Zhang, Yang and Felser, Claudia and Yan, Binghai},
  journal = {Phys. Rev. Lett.},
  volume = {117},
  issue = {14},
  pages = {146403},
  numpages = {5},
  year = {2016},
  month = {Sep},
  publisher = {American Physical Society},
  doi = {10.1103/PhysRevLett.117.146403},
  url = {https://link.aps.org/doi/10.1103/PhysRevLett.117.146403}
}

@article{RojasSanchez2013,
  title = {Spin pumping and inverse spin Hall effect in germanium},
  author = {Rojas-S\'anchez, J.-C. and Cubukcu, M. and Jain, A. and Vergnaud, C. and Portemont, C. and Ducruet, C. and Barski, A. and Marty, A. and Vila, L. and Attan\'e, J.-P. and Augendre, E. and Desfonds, G. and Gambarelli, S. and Jaffr\`es, H. and George, J.-M. and Jamet, M.},
  journal = {Phys. Rev. B},
  volume = {88},
  issue = {6},
  pages = {064403},
  numpages = {15},
  year = {2013},
  month = {Aug},
  publisher = {American Physical Society},
  doi = {10.1103/PhysRevB.88.064403}
}

@article{Zhang2024,
    author = {Zhang, Qin and Chen, Xu and Yu, Yue and Li, Huinan and Dou, Mingbo and Gurung, G. and Wang, Xianjie and Tao, L. L.},
    title = "{Spin Hall effect in doped ferroelectric HfO2}",
    journal = {Applied Physics Letters},
    volume = {125},
    number = {3},
    pages = {032905},
    year = {2024},
    month = {07},
    issn = {0003-6951},
    doi = {10.1063/5.0217628},
}

@article{Tao2017,
  title = {Reversible spin texture in ferroelectric $\mathrm{Hf}{\mathrm{O}}_{2}$},
  author = {Tao, L. L. and Paudel, Tula R. and Kovalev, Alexey A. and Tsymbal, Evgeny Y.},
  journal = {Phys. Rev. B},
  volume = {95},
  issue = {24},
  pages = {245141},
  numpages = {9},
  year = {2017},
  month = {Jun},
  publisher = {American Physical Society},
  doi = {10.1103/PhysRevB.95.245141},
  url = {https://link.aps.org/doi/10.1103/PhysRevB.95.245141}
}

@article{Guo2008,
  title = {Intrinsic Spin Hall Effect in Platinum: First-Principles Calculations},
  author = {Guo, G. Y. and Murakami, S. and Chen, T.-W. and Nagaosa, N.},
  journal = {Phys. Rev. Lett.},
  volume = {100},
  issue = {9},
  pages = {096401},
  numpages = {4},
  year = {2008},
  month = {Mar},
  publisher = {American Physical Society},
  doi = {10.1103/PhysRevLett.100.096401},
  url = {https://link.aps.org/doi/10.1103/PhysRevLett.100.096401}
}

@article{Ando2010,
    author = {Ando, K. and Saitoh, E.},
    title = "{Inverse spin-Hall effect in palladium at room temperature}",
    journal = {Journal of Applied Physics},
    volume = {108},
    number = {11},
    pages = {113925},
    year = {2010},
    month = {12},
    issn = {0021-8979},
    doi = {10.1063/1.3517131}
}

@article{Pai2012,
    author = {Pai, Chi-Feng and Liu, Luqiao and Li, Y. and Tseng, H. W. and Ralph, D. C. and Buhrman, R. A.},
    title = "{Spin transfer torque devices utilizing the giant spin Hall effect of tungsten}",
    journal = {Applied Physics Letters},
    volume = {101},
    number = {12},
    pages = {122404},
    year = {2012},
    month = {09},
    issn = {0003-6951},
    doi = {10.1063/1.4753947}
}

@article{Yu2018,
  title = {Determination of spin Hall angle and spin diffusion length in $\ensuremath{\beta}$-phase-dominated tantalum},
  author = {Yu, R. and Miao, B. F. and Sun, L. and Liu, Q. and Du, J. and Omelchenko, P. and Heinrich, B. and Wu, Mingzhong and Ding, H. F.},
  journal = {Phys. Rev. Mater.},
  volume = {2},
  issue = {7},
  pages = {074406},
  numpages = {7},
  year = {2018},
  month = {Jul},
  publisher = {American Physical Society},
  doi = {10.1103/PhysRevMaterials.2.074406},
  url = {https://link.aps.org/doi/10.1103/PhysRevMaterials.2.074406}
}

@article{Isasa2015,
  title = {Temperature dependence of spin diffusion length and spin Hall angle in Au and Pt},
  author = {Isasa, Miren and Villamor, Estitxu and Hueso, Luis E. and Gradhand, Martin and Casanova, F\`elix},
  journal = {Phys. Rev. B},
  volume = {91},
  issue = {2},
  pages = {024402},
  numpages = {7},
  year = {2015},
  month = {Jan},
  publisher = {American Physical Society},
  doi = {10.1103/PhysRevB.91.024402},
  url = {https://link.aps.org/doi/10.1103/PhysRevB.91.024402}
}

@article{Wang2014,
    author = {Wang, Yi and Deorani, Praveen and Qiu, Xuepeng and Kwon, Jae Hyun and Yang, Hyunsoo},
    title = "{Determination of intrinsic spin Hall angle in Pt}",
    journal = {Applied Physics Letters},
    volume = {105},
    number = {15},
    pages = {152412},
    year = {2014},
    month = {10},
    issn = {0003-6951},
    doi = {10.1063/1.4898593}}

@article{Zhang2020,
author = {Zhang, Wenxu and Teng, Zhao and Zeng, Huizhong and Zhang, Hongbin and Železný, Jakub and Zhang, Wanli},
title = {Tuning Spin Hall Conductivity in GeTe by Ferroelectric Polarization},
journal = {physica status solidi (b)},
volume = {257},
number = {9},
pages = {2000143},
doi = {https://doi.org/10.1002/pssb.202000143},
year = {2020}
}

@Article{Tsirkin2021,
author={Tsirkin, Stepan S.},
title={High performance Wannier interpolation of Berry curvature and related quantities with WannierBerri code},
journal={npj Computational Materials},
year={2021},
month={Feb},
day={19},
volume={7},
number={1},
pages={33},
issn={2057-3960},
doi={10.1038/s41524-021-00498-5},
url={https://doi.org/10.1038/s41524-021-00498-5}
}

@article{Pizzi2020,
doi = {10.1088/1361-648X/ab51ff},
url = {https://dx.doi.org/10.1088/1361-648X/ab51ff},
year = {2020},
month = {jan},
publisher = {IOP Publishing},
volume = {32},
number = {16},
pages = {165902},
author = {Giovanni Pizzi and Valerio Vitale and Ryotaro Arita and Stefan Blügel and Frank Freimuth and Guillaume Géranton and Marco Gibertini and Dominik Gresch and Charles Johnson and Takashi Koretsune and Julen Ibañez-Azpiroz and Hyungjun Lee and Jae-Mo Lihm and Daniel Marchand and Antimo Marrazzo and Yuriy Mokrousov and Jamal I Mustafa and Yoshiro Nohara and Yusuke Nomura and Lorenzo Paulatto and Samuel Poncé and Thomas Ponweiser and Junfeng Qiao and Florian Thöle and Stepan S Tsirkin and Małgorzata Wierzbowska and Nicola Marzari and David Vanderbilt and Ivo Souza and Arash A Mostofi and Jonathan R Yates},
title = {Wannier90 as a community code: new features and applications},
journal = {Journal of Physics: Condensed Matter}
}

@article{Perdew1996,
  title = {Generalized Gradient Approximation Made Simple},
  author = {Perdew, John P. and Burke, Kieron and Ernzerhof, Matthias},
  journal = {Phys. Rev. Lett.},
  volume = {77},
  issue = {18},
  pages = {3865--3868},
  numpages = {0},
  year = {1996},
  month = {Oct},
  publisher = {American Physical Society},
  doi = {10.1103/PhysRevLett.77.3865},
  url = {https://link.aps.org/doi/10.1103/PhysRevLett.77.3865}
}

@article{Giannozzi2017,
doi = {10.1088/1361-648X/aa8f79},
url = {https://dx.doi.org/10.1088/1361-648X/aa8f79},
year = {2017},
month = {oct},
publisher = {IOP Publishing},
volume = {29},
number = {46},
pages = {465901},
author = {P Giannozzi and O Andreussi and T Brumme and O Bunau and M Buongiorno Nardelli and M Calandra and R Car and C Cavazzoni and D Ceresoli and M Cococcioni and N Colonna and I Carnimeo and A Dal Corso and S de Gironcoli and P Delugas and R A DiStasio and A Ferretti and A Floris and G Fratesi and G Fugallo and R Gebauer and U Gerstmann and F Giustino and T Gorni and J Jia and M Kawamura and H-Y Ko and A Kokalj and E Küçükbenli and M Lazzeri and M Marsili and N Marzari and F Mauri and N L Nguyen and H-V Nguyen and A Otero-de-la-Roza and L Paulatto and S Poncé and D Rocca and R Sabatini and B Santra and M Schlipf and A P Seitsonen and A Smogunov and I Timrov and T Thonhauser and P Umari and N Vast and X Wu and S Baroni},
title = {Advanced capabilities for materials modelling with Quantum ESPRESSO},
journal = {Journal of Physics: Condensed Matter}
}
\end{document}